\documentclass[
reprint,
superscriptaddress,
amsmath,amssymb,
apl,
]{revtex4-2}

\usepackage[utf8]{inputenc}
\usepackage{graphicx,import}
\graphicspath{{Figures/}}
\usepackage{bm}
\usepackage[colorlinks]{hyperref}
\usepackage[all]{hypcap}
\usepackage{xcolor}
\usepackage{braket}
\usepackage{siunitx}
\usepackage{changes}
\usepackage{lipsum}

\usepackage{svg} 

\definecolor{bluegray}{RGB}{40,180,160}
\definecolor{navygray}{RGB}{110,140,170}
\definecolor{meadowgreen}{RGB}{0,128,0}
\definecolor{coolbrown}{RGB} {165,42,42}

\DeclareSIUnit{\sq}{\Box}

\usepackage[capitalise,nameinlink]{cleveref} 
\newcommand{\crefadd}[2]{\hyperref[#1]{\cref{#1}\,#2}}
\newcommand{\Crefadd}[2]{\hyperref[#1]{\Cref{#1}\,#2}}

\newcommand{\fq}{f_\mathrm{q}}

\hypersetup{
citecolor={navygray}, 
linkcolor={navygray},
urlcolor={navygray},
}

\makeatletter
\newcommand*{\balancecolsandclearpage}{%
  \close@column@grid
  \cleardoublepage
  \twocolumngrid
}
\makeatother

\begin{document}

\title{Spin Environment of a Superconducting Qubit in High Magnetic Fields}

\author{S.~Günzler}
\email{simon.guenzler@kit.edu}
\affiliation{PHI,~Karlsruhe~Institute~of~Technology,~76131~Karlsruhe,~Germany}
\affiliation{IQMT,~Karlsruhe~Institute~of~Technology,~76131~Karlsruhe,~Germany}

\author{J.~Beck}
\thanks{First two authors contributed equally.}
\affiliation{PHI,~Karlsruhe~Institute~of~Technology,~76131~Karlsruhe,~Germany}

\author{D.~Rieger}
\affiliation{PHI,~Karlsruhe~Institute~of~Technology,~76131~Karlsruhe,~Germany}

\author{N.~Gosling}
\affiliation{IQMT,~Karlsruhe~Institute~of~Technology,~76131~Karlsruhe,~Germany}

\author{N.~Zapata}
\affiliation{IQMT,~Karlsruhe~Institute~of~Technology,~76131~Karlsruhe,~Germany}

\author{M.~Field}
\affiliation{IQMT,~Karlsruhe~Institute~of~Technology,~76131~Karlsruhe,~Germany}

\author{S.~Geisert}
\affiliation{IQMT,~Karlsruhe~Institute~of~Technology,~76131~Karlsruhe,~Germany}

\author{A.~Bacher}
\affiliation{IMT,~Karlsruhe~Institute~of~Technology,~76131~Karlsruhe,~Germany}
\affiliation{KNMFi,~Karlsruhe~Institute~of~Technology,~76131~Karlsruhe,~Germany}

\author{J.~K.~Hohmann}
\affiliation{IMT,~Karlsruhe~Institute~of~Technology,~76131~Karlsruhe,~Germany}
\affiliation{KNMFi,~Karlsruhe~Institute~of~Technology,~76131~Karlsruhe,~Germany}

\author{M.~Spiecker}
\affiliation{PHI,~Karlsruhe~Institute~of~Technology,~76131~Karlsruhe,~Germany}
\affiliation{IQMT,~Karlsruhe~Institute~of~Technology,~76131~Karlsruhe,~Germany}

\author{W.~Wernsdorfer}
\affiliation{PHI,~Karlsruhe~Institute~of~Technology,~76131~Karlsruhe,~Germany}
\affiliation{IQMT,~Karlsruhe~Institute~of~Technology,~76131~Karlsruhe,~Germany}

\author{I.~M.~Pop}
\email{ioan.pop@kit.edu}
\affiliation{PHI,~Karlsruhe~Institute~of~Technology,~76131~Karlsruhe,~Germany}
\affiliation{IQMT,~Karlsruhe~Institute~of~Technology,~76131~Karlsruhe,~Germany}
\affiliation{Physics~Institute~1,~Stuttgart~University,~70569~Stuttgart,~Germany}

\begin{abstract}
Superconducting qubits equipped with quantum non-demolition readout and active feedback can be used as information engines to probe and manipulate microscopic degrees of freedom, whether intentionally designed or naturally occurring in their environment. 
In the case of spin systems, the required magnetic field bias presents a challenge for superconductors and Josephson junctions. 
Here we demonstrate a granular aluminum nanojunction fluxonium qubit (gralmonium) with spectrum and coherence resilient to fields beyond one Tesla. 
Sweeping the field reveals a paramagnetic spin\mbox{-}1/2 ensemble, which is the dominant gralmonium loss mechanism when the electron spin resonance matches the qubit. 
We also observe a suppression of \unit{\mega\hertz} range fast flux noise in magnetic field, suggesting the freezing of surface spins. 
Using an active state stabilization sequence, the qubit hyperpolarizes long-lived two-level systems (TLSs) in its environment, previously speculated to be spins. 
Surprisingly, the coupling to these TLSs is unaffected by magnetic fields, leaving the question of their origin open.
The robust operation of gralmoniums in Tesla fields offers new opportunities to explore unresolved questions in spin environment dynamics and facilitates hybrid architectures linking superconducting qubits with spin systems.
\end{abstract}

\maketitle
Superconducting qubits have rapidly evolved from proof-of-concept demonstrations to precision-engineered devices within the cQED framework~\cite{Blais2021May}, featuring quantum non-demolition readout and real-time feedback. 
These advances have enabled the observation of quantum jumps and trajectories~\cite{Vijay2011Mar, Ficheux2018May, Minev2019Jun}, active feedback error correction~\cite{Vijay2012Oct, Sivak2023Apr, Krinner2022May} and the exploration of quantum mechanics foundations~\cite{Saxberg2022Dec, Bild2023Apr, Storz2023May}.
Such precise control renders superconducting circuits ideal for interfacing with other mesoscopic degrees of freedom (DOFs), which may be deliberately integrated into hybrid architectures or arise from spurious microscopic systems that impair qubit performance.
Hybrid quantum architectures, where superconducting circuits couple to less amenable but longer-lived, magnetic-field-sensitive DOFs, have already demonstrated impressive achievements, such as
coherent spin-photon interactions~\cite{Landig2018Aug,Samkharadze2018Jan, Mi2018Mar}, spin ensemble~\cite{Bienfait2016Mar, Eichler2017Jan,Bienfait2016_2}
and even single-spin detection~\cite{Wang2023Jul, OSullivan2024Oct} using superconducting resonators, as well as single-magnon detection with a superconducting qubit~\cite{Lachance-Quirion2020Jan}.
Concurrently, various spurious environmental DOFs with often unknown magnetic field susceptibility are pervasive in superconducting devices.
These include quasiparticles~\cite{Serniak2019Jul, McEwen2024Feb, Connolly2024May, Krause2024Mar}, charge offsets~\cite{Serniak2019Jul, Christensen2019Oct}, spins~\cite{Kumar2016Oct, Anton2013Apr, Braumuller2020May, Stern2014Sep, Yoshihara2006Oct, Nugroho2013Apr, Yan2016Nov} and other TLS environments~\cite{Muller2019Oct, Thorbeck2023Jun, Spiecker2023Sep, Odeh2023Dec}.

High magnetic fields offer a powerful tool to characterize and tune various DOFs coupled to superconducting qubits, yet they are rarely utilized. 
This is explained by the fragility of aluminum-based devices in magnetic fields, as the superconducting gap is suppressed at $\sim\qty{10}{\milli\tesla}$ in bulk, and the Josephson junction (JJ) critical current diminishes in a Fraunhofer pattern. 
Utilizing thin aluminum films can improve field compatibility~\cite{Winkel2020Aug, Krause2022Mar, Rower2023May}, nevertheless, it still entails significant suppression of the qubit spectrum and coherence in the range of few hundred \unit{\milli\tesla}.
While the reduction of the superconducting gap can be mitigated by using field-resilient, low-loss superconductors like Nb~\cite{OSullivan2024Oct}, granular Aluminum (grAl)~\cite{Borisov2020Sep} or NbTiN~\cite{Samkharadze2016Apr,Kroll2019Jun}, finding a source of nonlinearity that maintains resilience under magnetic fields is considerably more challenging.
Efforts to develop field-resilient JJs that avoid Fraunhofer interference patterns include gate-tunable JJs based on semiconducting nanowires~\cite{Kringhoj2021May, Pita-Vidal2020Dec} or graphene layers ~\cite{Kroll2018Nov}. 
However, these JJs have shown marginal coherence, with qubit spectra exhibiting significant instability.

We overcome these limitations by using a granular aluminum nanojunction fluxonium qubit, known as gralmonium. 
This qubit combines the grAl field resilience~\cite{Deutscher1977Nov}
with the unique benefits of the grAl nanojunction~\cite{gralmonium}: 
low microwave losses and a compact nanoscopic footprint that eliminates Fraunhofer interference. 
We measure energy decay times $T_1\approx \qty{8}{\micro\second}$ and coherence times $ T_{2\mathrm{E}}\approx \qty{3}{\micro\second}$, robust in fields beyond \qty{1}{\tesla}, with less than \qty{2}{\percent} qubit frequency shift in this entire range.
We identify a paramagnetic spin ensemble coupled to the gralmonium, showcasing its potential for sensing. 
We also observe a decrease in the fast flux noise in Hahn echo experiments in magnetic field, indicating a freezing of the spin ensemble above \qty{400}{\milli\tesla}.
\begin{figure*}[t!]
\centering
\includegraphics[width=\textwidth]{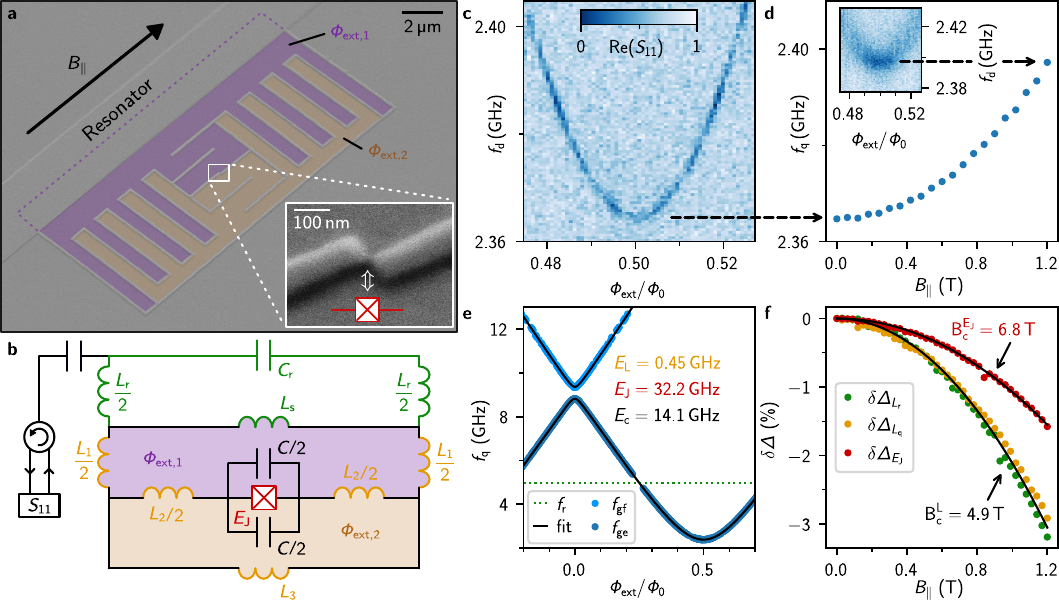}
\caption{\textbf{Gradiometric gralmonium qubit resilient to tesla magnetic field.} 
\textbf{a}~False-colored scanning electron microscope (SEM) image of the qubit circuit, galvanically coupled to the readout resonator. 
The device consists of a \qty{20}{\nano\meter} thick single layer of grAl. 
The colored regions (ocher \& violet) illustrate the $10\%$ mismatched areas of the two flux loops in the gradiometric design~\cite{Gusenkova2022Jan}, which  result in an effective flux bias $\Phi_\text{ext}$ in perpendicular magnetic field $B_\perp$ (cf.~\cref{eq:phiExtTotal}).
Inset: zoom-in on the $\sim \qty{20}{\nano\meter}$ wide grAl nanojunction of the qubit (cf.~\cite{gralmonium}).
\textbf{b}~Circuit schematic for the gradiometric qubit depicted in \textbf{a}:
the nanojunction (red) is shunted by an interdigitated capacitor and two flux loops  (ocher \& violet) with inductances $L_1 + L_\mathrm{s}$ and $L_3$, respectively. The inductance shared between the loops is $L_2$. 
The qubit is inductively coupled via $L_\mathrm{s}$ to the readout resonator (inductance $L_\mathrm{r}$, capacitance $C_\mathrm{r}$) for which we measure the  single-port reflection coefficient $S_{11}$.
\textbf{c}~Two-tone (TT) spectroscopy at the half flux sweet spot $\Phi_\textbf{ext} = \Phi_0/2$ in $B_\parallel = \qty{0}{\tesla}$.
\textbf{d}~Increase of the sweet spot qubit frequency in magnetic field up to \qty{1,2}{\tesla}.
Inset: TT-spectroscopy in $B_\parallel = \qty{1,2}{\tesla}$.
\textbf{e}~Qubit spectrum: ground to excited ($f_{\text{ge}}$ in dark blue markers) and ground to second-excited ($f_{\text{gf}}$ in light blue markers) state transitions extracted from TT-spectroscopy. 
From a fit (black line) to the fluxonium Hamiltonian  (\cref{eq:fluxoniumHamiltonian}), we obtain $E_\mathrm{J}/h = \qty{32.2}{\giga\hertz}$ (i.e. critical current $I_\mathrm{c}=\qty{64.9}{\nano\ampere}$),  $E_\mathrm{c}/h= \qty{14.1}{\giga\hertz}$ ($C=\qty{1.37}{\femto\farad}$) and $E_\mathrm{L}/h = \qty{0.454}{\giga\hertz}$ ($L_\mathrm{q}=\qty{360}{\nano\henry}$).
\textbf{f}~Suppression of the grAl superconducting gap $\Delta$ in magnetic field.
The red and orange markers, corresponding to the qubit nanojunction and inductor superconducting gaps ($\Delta_{E_\text{J}}$, $\Delta_{L_\mathrm{q}}$), are obtained from fitted $E_\text{J}$ and $E_\text{L}$ values (cf.~\textbf{e}) at each magnetic field. The capacitance $C$ is fixed to the fit value obtained in  $B_\parallel =\qty{0}{\tesla}$.
The green markers are obtained from the shift of the readout resonator frequency $f_{\text{r}}(B_\parallel)$.
The black lines show fits to the field dependence of the superconducting gap, indicating a $40\%$ higher critical field for the nanojunction.
}
\label{fig:sample}
\end{figure*}
Moreover, we find the qubit to be coupled to a recently discovered, long-lived TLS ensemble~\cite{Spiecker2023Sep, Zhuang2025Mar}, which accounts for half of the dissipation budget.
Notably, we do not observe a magnetic dependence of this coupling, challenging the recently proposed spins hypothesis as its origin~\cite{Spiecker2023Sep}. 
Finally, we show that the critical current noise reported in Ref~\cite{gralmonium} is not magnetic field susceptible.

In \cref{fig:sample} we present the field resilient gralmonium qubit,
fabricated from a single layer of grAl (cf.~\crefadd{fig:sample}{a}), with a critical field on the order of $B_\mathrm{c}\sim\qty{6}{\tesla}$~\cite{Borisov2020Sep}.
We use a \qty{20}{\nano\metre} thick grAl film with a sheet inductance of $L_\square = \qty{0,75}{\nano\henry/\sq}$ (resistivity $\rho = \qty{2000}{\micro\ohm\centi\metre}$) to design all circuit elements (cf.~\cref{sec:supp:fabrication}).
We galvanically couple a \qty{1}{\milli\metre} long stripline readout resonator to the qubit circuit, consisting of a superinductor, a geometric finger capacitance and a grAl nanojunction.
Implemented by a $\sim(\qty{20}{\nano\metre})^3$ grAl volume, the nanojunction offers a sinusoidal current-phase relation similar to conventional Al/AlO$_x$/Al JJ~\cite{gralmonium}, while exposing a minute cross-section to Fraunhofer interference.
To reduce the sensitivity to magnetic flux fluctuations perpendicular to the thin film, we implement a gradiometric design~\cite{Gusenkova2022Jan} with two flux loops (ocher \& violet in~\crefadd{fig:sample}{a}) containing fluxes $\Phi_1, \Phi_2$, respectively.
The equivalent circuit diagram in \crefadd{fig:sample}{b} can be mapped to the standard fluxonium Hamiltonian
\begin{align}
    H &= 4 E_\text{C} \hat{n}^2 + \frac{1}{2}E_\text{L}\left(\hat{\varphi} - 2\pi\frac{\Phi_\text{ext}}{\Phi_0}\right)^2 - E_\text{J}\cos\hat{\varphi}\,,
    \label{eq:fluxoniumHamiltonian}
\end{align}
where $E_\text{L}=(\Phi_0 / 2\pi)^2 / L_\text{q}$, $E_\text{C}=e^2 / 2 C$, $E_\text{J}=I_\text{c} \Phi_0 /2\pi $ and $\Phi_0=\text{h}/2\text{e}$. 
Here, $\hat{n}$ represents the number of Cooper pairs and $\hat{\varphi}$ is the phase difference across the junction.
Due to the low intrinsic capacitance of the nanojunction, the qubit charging energy $E_\text{C}$ is dominated by the interdigitated capacitor $C$~\cite{gralmonium}.
For the gradiometric circuit, the effective qubit inductance  is given by $L_\text{q}=\frac{L_{1,\mathrm{s}}L_2+L_2 L_3+L_3L_{1,\mathrm{s}}}{L_{1,\mathrm{s}}+L_3}$ with $L_{1,\mathrm{s}}=L_1+L_\mathrm{s}$, and the effective external flux is
\begin{align}
    & \Phi_{\text{ext}} = \Phi_\Delta - \alpha \Phi_\Sigma  \,.
    \label{eq:phiExtTotal}
\end{align}
Here, $\Phi_{\Sigma/\Delta} =  \frac{\Phi_\text{ext,1}}{2} \pm \frac{\Phi_\text{ext,2}}{2}$ denote the mean and difference of fluxes in the two loops, respectively, and $\alpha = \frac{L_{1,\mathrm{s}}-L_3}{L_{1,\mathrm{s}}+L_3} $ is the inductance asymmetry.
In our gradiometric design, the magnetic flux susceptibility is reduced by a factor of $\Phi_{\text{ext},1}/\Phi_\Delta = 4.6$ with $\alpha\approx 0$ (cf. \cref{sec:supp:fluxsweep_comparison}).

From two-tone (TT) spectroscopy at half flux bias $\Phi=\Phi_0/2$ shown in ~\crefadd{fig:sample}{c}, we determine a qubit frequency of $\fq (\Phi_0/2)=\qty{2,365}{\giga\hertz}$ in zero field, $B_\parallel = 0$.
Remarkably, as shown in \crefadd{fig:sample}{d}, tracking the sweet spot qubit frequency in magnetic field reveals only a $1\%$ increase (\qty{32}{\mega\hertz}) up to \qty{1,2}{\tesla}, illustrating the compatibility of the gradiometric gralmonium qubit with high magnetic fields.
The spectroscopy data in \qty{1,2}{\tesla} is blurred compared to zero field due to low-frequency flux noise, likely from vibrations of the sample holder inside the vector magnet (cf.~\cref{sec:supp:sampleholder}).
\Crefadd{fig:sample}{e} shows the gralmonium spectrum up to \qty{13}{\giga\hertz}, extracted from two-tone spectroscopy.
A joint fit of the qubit transitions $\ket{\text{g}}\to\ket{\text{e}}$ and $\ket{\text{g}}\to\ket{\text{f}}$ to a numerical diagonalization of \cref{eq:fluxoniumHamiltonian} yields typical fluxonium parameters: $E_\mathrm{J}/h = \qty{32.2}{\giga\hertz}$,  $E_\mathrm{c}/h = \qty{14.1}{\giga\hertz}$ and $E_\mathrm{L}/h = \qty{0.454}{\giga\hertz}$.

To assess the effect of magnetic field on the fluxonium parameters, we measure the qubit ground to excited transition frequency $f_\mathrm{ge}$ near the half- and zeroflux sweet spots at each $B_\parallel$, using two-tone spectroscopy (similar to ~\cref{fig:sample}{c}).
We fit $f_\mathrm{ge}$ to~\cref{eq:fluxoniumHamiltonian} using the field independent capacitance $C=\qty{1,37}{\femto\farad}$ obtained in zero field.
From the fitted parameters, using $E_\text{J} \propto \Delta(B_\parallel)$ and $L_{\text{kin}}\propto 1/\Delta(B_\parallel) $, we extract the magnetic field suppression of the superconducting gap in the superinductor ($\delta \Delta_{L_\mathrm{q}}$) and nanojunction ($\delta \Delta_{E_\text{J}}$), as shown in~\crefadd{fig:sample}{f}. 
The suppression of the resonator superconducting gap ($\delta \Delta_{L_\mathrm{r}}$) is independently obtained by fitting the resonance frequency shift $f_\text{r} \propto 1/\sqrt{L_{\text{kin}}}$.
Interestingly, the nanojunction has an even higher field resilience than the grAl resonator and superinductor.
We fit the relative gap suppression to $\sqrt{1-(B_\parallel/B_\mathrm{c})^2}$~\cite{Douglass1961Apr} and obtain  a critical field $B_\mathrm{c}^{E_\mathrm{J}} = \qty{6,8}{\tesla}$ for the nanojunction and $B_\mathrm{c}^\mathrm{L} = \qty{4,9}{\tesla}$ for the resonator and qubit inductance.
The fact that $B_\mathrm{c}^{E_\mathrm{J}}>B_\mathrm{c}^\mathrm{L}$ indicates that possible Fraunhofer interference in the nanojunction plays a minor role.
The higher critical field of the nanojunction is not understood and could be due to its reduced dimensions, similar to Ref.~\cite{Deshpande2024Aug}.
\begin{figure*}[t!]
\centering
\includegraphics[width=\textwidth]{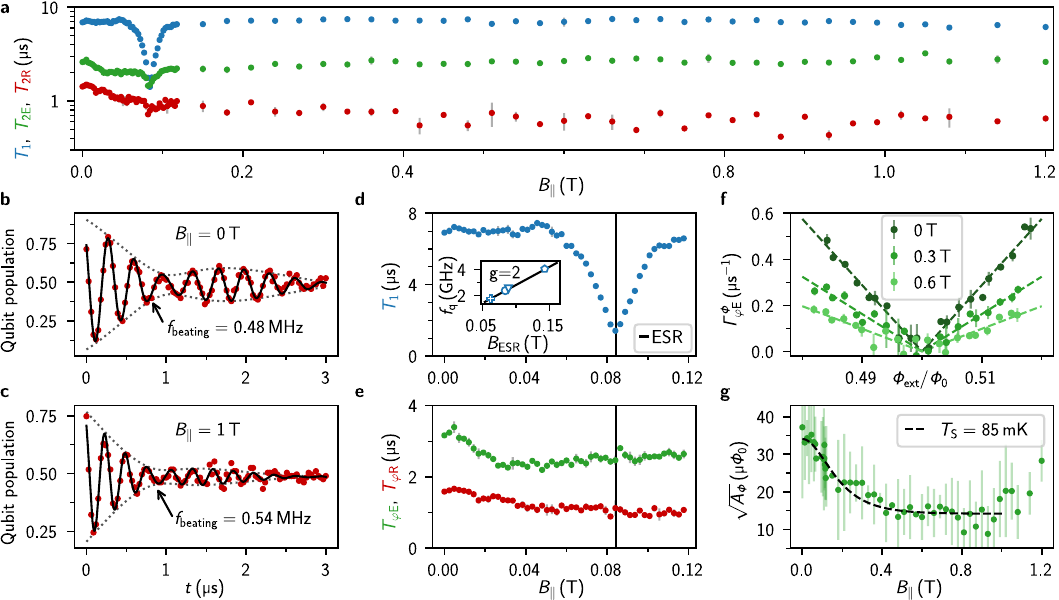}
\caption{
\textbf{Qubit coherence in magnetic field: signatures of environmental spin polarization.} 
\textbf{a}~Energy relaxation time~$T_1$, Ramsey and echo coherence time, $T_{2 \mathrm{R}}$ and $T_{2 \mathrm{E}}$ respectively, of the gradiometric gralmonium in magnetic field up to~\qty{1}{\tesla}. 
\textbf{b, c}~Ramsey fringes measured in $B_\parallel = \qty{0}{\tesla}$ and $B_\parallel = \qty{1}{\tesla}$, respectively.
A two-frequency fit (black line) indicates a similar beating pattern (dotted envelope) for both magnetic fields.
\textbf{d}~Energy relaxation~$T_1$ up to~\qty{120}{\milli\tesla}:
similarly to observations on resonators~\cite{Samkharadze2016Apr, Kroll2019Jun, Borisov2020Sep}, the drop in $T_1$ suggests  coupling to the electron spin resonance (ESR) of paramagnetic impurities of unknown origin.
Inset: The fields $B_\text{ESR} = h \fq / g \mu_\mathrm{B} $ at which the ESR matches different qubit frequencies in different cooldowns, correspond to the expectation for a spin $s=1/2$ ensemble with gyromagnetic factor $g=2$ (black line). 
Note that we use  the same device  for which the qubit frequency changes between cooldowns (cf. Ref.~\cite{gralmonium}).
\textbf{e} Dephasing times~$T_{\varphi \mathrm{R}}$,~$T_{\varphi \mathrm{E}}$ remain unaffected by the ESR.
\textbf{f}~Flux noise echo dephasing rate $\Gamma_{\varphi \text{E}}^\Phi$ in the vicinity of the sweet spot for three in-plane magnetic fields. 
Dashed lines show fits to \cref{eq:GammaE_flux}.
\textbf{g}~Flux noise amplitude $\sqrt{A_\Phi}$ in magnetic field with fit to \cref{eq:APhi_vs_B}, corresponding to a spin freezing with a spin temperature of $T_\mathrm{S}=\qty{85}{\milli\kelvin}$.
In all panels, the errorbars represent the standard deviation obtained from successive measurements.
} 
\label{fig:coherence}
\end{figure*}
\begin{figure*}[t!]
\centering
\includegraphics[width=\textwidth]{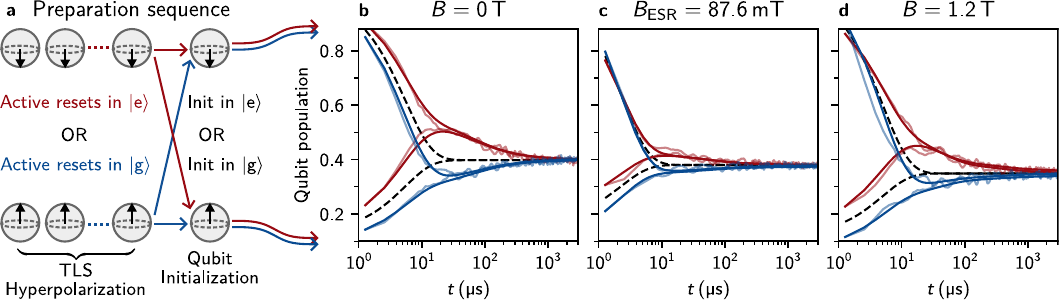}
\caption{
\textbf{Magnetic susceptibility of long-lived two-level-systems (TLSs) in high field.}
\textbf{a}~Sketch of the qubit preparation sequence used in panels (\textbf{b-d}). 
The repeated ($N=10^4$) active reset of the qubit state in $\ket{\text{g}}$ or $\ket{\text{e}}$ (blue and red traces in all panels, respectively) results in the hyperpolarization of environmental, long-lived TLS~\cite{Spiecker2023Sep}.
The last step of the preparation sequence consists in a qubit initialization in $\ket{\text{g}}$ or $\ket{\text{e}}$. 
We use a \qty{540}{\nano\second} rectangular readout pulse and a \qty{32}{\nano\second} Gaussian $\pi$-pulse. 
\textbf{b}~Qubit population relaxation after the preparation sequence for different magnetic fields $B_\parallel$.
We fit the data (semi-transparent) to the theoretical model \cite{Spiecker2023Sep, Spiecker2024May} (opaque).
For reference, the black dashed lines show an exponential decay with the qubit energy relaxation rate $\Gamma_1$.
In zero field, we reproduce the signatures of TLS hyperpolarization recently observed in other superconducting qubits \cite{Spiecker2023Sep, Odeh2023Dec}, i.e. undershoot (blue) and overshoot (red) compared to the single exponential decay.
\textbf{c}~At the ESR resonance field $B_\text{ESR}$, the hyperpolarization signatures are suppressed due to energy relaxation of the qubit into the paramagnetic ensemble.
\textbf{d}~The signatures of TLS hyperpolarization on qubit relaxation in magnetic fields exceeding \qty{1}{T} are comparable to zero field, indicating a very low susceptibility of the long-lived TLSs to magnetic field.
}
\label{fig:hyperpolarization}
\end{figure*}

We quantify the quantum coherence of the gralmonium in field by performing time-domain measurements at the half flux sweet spot, as summarized in~\cref{fig:coherence}.
Remarkably, the energy relaxation time $T_1$ and Hahn echo coherence time $T_{2\mathrm{E}}$ remain robust in fields up to \qty{1,2}{\tesla}, the upper limit of our vector magnet (cf.~\crefadd{fig:coherence}{a}).
The Ramsey coherence time $T_{2\mathrm{R}}$ decreases from a maximum of $T_{2\mathrm{R}}=\qty{1,5}{\micro\second}$ to $T_{2\mathrm{R}}=\qty{0,7}{\micro\second}$ in fields above \qty{1}{\tesla}.
We attribute this to an increase of low frequency flux noise, which stems from global flux variations introduced by vibrations and activated vortices in the vector magnet~\cite{Schwarz2013Jan}, or from local flux noise, possibly from spins clusters~\cite{Rower2023May}.

The grAl nanojunction exhibits critical current fluctuations, as evidenced by a \qty{0,5}{\mega\hertz} toggling of the qubit frequency and a corresponding beating pattern in Ramsey fringes at zero field (cf.~\crefadd{fig:coherence}{b}).
As demonstrated in Ref.~\cite{gralmonium}, these fluctuations are inconsistent with transverse coupling to a fixed frequency TLS, but originate from fluctuations of the nanojunction energy, potentially arising from structural defects, charge noise, or paramagnetic impurities.
This issue is also relevant for standard Al/AlO$_x$/Al tunnel JJs~\cite{Nugroho2013Apr,VanHarlingen2004Aug}.
We leverage the gradiometric gralmonium’s field resilience to test the magnetic susceptibility of critical current noise, showing in \crefadd{fig:coherence}{c} that a \qty{1}{\tesla} in-plane magnetic field does not suppress the discrete fluctuations of the Josephson energy. 
This observation excludes magnetically susceptible sources, such as a local spin environment, as the origin of these fluctuations. 
Further experiments, such as electric field bias or mechanical strain on the substrate~\cite{Muller2019Oct, Kristen2024May} or spin-locking TLS spectroscopy~\cite{Abdurakhimov2022Dec}, are required to identify their cause.

We observe a dip in the energy relaxation time at the magnetic field where the electron spin resonance (ESR) matches the qubit frequency $h f = g \mu_\text{B} B$ (cf.~\crefadd{fig:coherence}{d}).
This ESR resonance does not impact the dephasing times $T_{\varphi \mathrm{R}}$, $T_{\varphi \mathrm{E}}$ (cf.~\crefadd{fig:coherence}{e}), as expected in the limit of a coupling strength much smaller than the qubit linewidth~\cite{Spiecker2024May}.
By exploiting $\Delta E_\mathrm{J}\sim\unit{\giga\hertz}$ changes in the nanojunction energy after thermal cycling, we change the qubit frequency of the same device, allowing measurements of the ESR-resonant field $B_\text{ESR}$ across multiple qubit frequencies at the half-flux sweet spot (inset of~\crefadd{fig:coherence}{d}). 
The linear trend of the extracted magnetic fields $B_\text{ESR}$ aligns with the prediction for a $g=2$ spin $s=1/2$ ensemble, identifying a paramagnetic spin ensemble coupled to our qubit.

\Crefadd{fig:coherence}{f} illustrates the flux dependence of the Hahn echo flux noise dephasing rate, $\Gamma_{\varphi \text{E}}^\Phi$, near the sweet spot for three in-plane magnetic fields ($B_\parallel = 0$, $B_\parallel = \qty{0,3}{\tesla}$, $B_\parallel = \qty{0,6}{\tesla}$).
Away from the sweet spot, we observe a Gaussian contribution in the Hahn echo decay curve, consistent with commonly observed $1/f$ flux noise~\cite{Rower2023May,Stern2014Sep, Yoshihara2006Oct, Braumuller2020May, Yan2016Nov, Nugroho2013Apr, Schriefl2006Jan}.
We extract $\Gamma_{\varphi \text{E}}^\Phi$ from the flux-dependent Gaussian envelope  $e^{-(\Gamma_{\varphi \text{E}}^\Phi t)^2}$, on top of a purely exponential decay, $e^{-(\Gamma_1/2 + \Gamma_{\varphi\text{E}}^\text{const})t}$, extracted at $\Phi=\Phi_0/2$.
The flux-independent term $\Gamma_{\varphi \text{E}}^\text{const}$ may originate from critical current noise or photon shot noise; in the latter case the residual photon number is $\bar{n}=0.27$, corresponding to an effective temperature of $\qty{150}{\milli\kelvin}$, in agreement with the qubit temperature (cf.~\cref{sec:supp:sampleholder}).
Interestingly, the flux dependence $\Gamma_{\varphi \text{E}}^\Phi(\Phi_\text{ext})$ weakens as the magnetic field $B_\parallel$ increases (cf.~\crefadd{fig:coherence}{f}), reminiscent of earlier observations in flux qubits at lower field~\cite{Rower2023May}.
We fit the flux noise amplitude $\sqrt{A_\Phi }$ for a $S_\Phi(\omega) = A_\Phi/\omega$  power spectral density using~\cite{Yoshihara2006Oct, Schriefl2006Jan}
\begin{align}
    \Gamma_{\varphi \text{E}}^\Phi=\sqrt{A_\Phi \ln 2}\left|\frac{\partial \omega}{\partial \Phi_\text{ext}}\right| \,.
    \label{eq:GammaE_flux}
\end{align}
With increasing $B_\parallel$, we observe a decrease of $\sqrt{A_\Phi }$ by a factor of $\sim 2$, reported in \crefadd{fig:coherence}{g}, which holds across different qubit frequencies in several cooldowns (cf.~\cref{sec:supp:fluxNoise}).
However, for $B_\parallel\gtrsim \qty{1}{\tesla}$,  $\sqrt{A_\Phi }$ increases, suggesting the onset of a competing mechanism, likely due to vortex dynamics in the magnet wires.

We model the flux noise as the sum of a large number of magnetic two-level fluctuators, consistent with the commonly accepted spin-based origin of flux noise~\cite{Lanting2009Feb, Quintana2017Jan, Gao2025Apr}.
Each of them constitutes a source of asymmetric random telegraphic noise, with a Lorentzian power spectrum $S(\omega)\propto(\Gamma_1/\Gamma_\uparrow + \Gamma_1/\Gamma_\downarrow)^{-1} \cdot \Gamma_1/(\Gamma_1^2 + \omega^2) $, where $\Gamma_1=\Gamma_\uparrow + \Gamma_\downarrow$ are the excitation and relaxation rates of the fluctuator, respectively~\cite{Kogan1996Aug}. 
In the limit of identical fluctuators, $S(\omega)$ remains Lorentzian, while for fluctuators with $1/\Gamma_1$ uniformly distributed, $S(\omega)\propto 1/\omega$~\cite{Schriefl2006Jan}.
However, for any distribution, the amplitude of the power spectrum is $A_\Phi\propto (\Gamma_1/\Gamma_\uparrow + \Gamma_1/\Gamma_\downarrow)^{-1}$, which becomes (cf.~\cref{sec:supp:fluxNoise})
\begin{align}
    A_\Phi\propto  1/\cosh^2\left(\frac{\mu_\mathrm{B} B}{k_\mathrm{B}T_\mathrm{S}} \right)\,. \label{eq:APhi_vs_B}
\end{align}
Here, $2\mu_\mathrm{B}B$ is the energy of $g=2, s=1/2$ paramagnetic impurities and $\mu_\mathrm{B}$, $k_\mathrm{B}$ and $T_S$ are the Bohr magneton, the Boltzmann constant and the spin bath temperature, respectively.
A fit with $T_\mathrm{S}=\qty{85}{\milli\kelvin}$ aligns with the measured flux noise amplitude (black line in \crefadd{fig:coherence}{g}) up to \qty{1}{\tesla}.
This suggests the freezing of $g=2$ paramagnetic impurities responsible for the reduction of flux noise, presumably the same spin environment that causes the $T_1$ dip (cf.~\crefadd{fig:coherence}{d}).

In \cref{fig:hyperpolarization}, we leverage the field resilience of the gralmonium to probe the magnetic susceptibility of a recently discovered TLS bath coupled to superconducting qubits~\cite{Spiecker2023Sep, Zhuang2025Mar, Spiecker2024May}. 
These TLSs have been shown to induce non-Markovian qubit dynamics and their long lifetime, exceeding $1/\Gamma_\text{TLS}\geq \qty{50}{\milli\second}$, makes a spin-based origin plausible.
Following Ref.~\cite{Spiecker2023Sep}, by repeatedly preparing the qubit in either $\ket{\text{g}}$ or $\ket{\text{e}}$ using fast feedback over $N=10^4$ iterations, the TLS ensemble hyperpolarizes via its cross-relaxation to the qubit.
After this polarization sequence, the qubit is initialized in either $\ket{\text{g}}$ or $\ket{\text{e}}$, and its population is monitored using stroboscopic quantum jump measurements.
\Crefadd{fig:hyperpolarization}{b} shows the distinct signatures of a hyperpolarized long-lived TLS ensemble coupled to the gralmonium:
regardless of the qubit's initial state, it relaxes to the TLS ensemble population on a $T_1$ timescale, while the TLS ensemble itself decays to thermal equilibrium on milliseconds timescale.
By modeling the qubit coupled to a ladder of $10^2$ TLSs~\cite{Spiecker2023Sep}, we extract a gralmonium relaxation $\Gamma_1 = 1/\qty{5,4}{\micro\second}$, of which TLS cross-relaxation accounts for $\sum_k\Gamma_\text{qt}^k = 1/\qty{22}{\micro\second}$.

In magnetic field, the signatures of TLS hyperpolarization remain visible, as illustrated in \crefadd{fig:hyperpolarization}{c,d}.
The fact that the hyperpolarization in $B_\parallel=\qty{1,2}{\tesla}$  is comparable to zero field indicates that the TLS bath is not magnetically susceptible, ruling out origins such as electronic spins.
Remaining non-magnetically-susceptible microscopic origins include subgap states, possibly  trapped quasiparticles~\cite{deGraaf2020Dec}.
As shown in ~\crefadd{fig:hyperpolarization}{c}, at $B\approx B_\text{ESR}$, where $T_1$ is suppressed by a factor of $7$ (cf.~\crefadd{fig:coherence}{d}), the TLS hyperpolarization is less pronounced.
Therefore, we are still able to hyperpolarize the long lived TLSs, but not the paramagnetic spins.
This indicates that the spin ensemble is large enough or sufficiently coupled to the environment that it embodies a Markovian bath.
In contrast, the long-lived TLS environment appears to be uncoupled to the spin ensemble, as evidenced by the fit in~\crefadd{fig:hyperpolarization}{c} with a practically unchanged cross-relaxation rate of $\sum_k\Gamma_\text{qt}^k = 1/\qty{33}{\micro\second}$.

In conclusion, we have introduced a field-resilient superconducting qubit — the gradiometric gralmonium — that operates robustly in Tesla magnetic fields.
By incorporating a grAl nanojunction, the gralmonium maintains spectral stability and coherence in high magnetic fields, circumventing the Fraunhofer interference typically observed in Josephson junction-based superconducting circuits.
We reveal distinct properties of spin environments coupled to the gralmonium by addressing their magnetic field susceptibility. 
Using ESR, we characterize a paramagnetic spin-1/2 ensemble that couples transversely to the qubit, demonstrating the gralmonium’s potential as a probe for spin dynamics. 
We confirm the long-standing hypothesis of the freeze-out of fast flux noise in high fields, consistent with a spin $s=1/2$, $g=2$ paramagnetic origin.
The operation of the gralmonium in magnetic field also allowed us to disprove the electron-spin hypothesis for the long-lived two-level system (TLS) environment responsible for non-Markovian qubit dynamics.

Future work should address flux noise suppression and spectral noise analysis, and should validate the correlation between flux noise and the spin-$1/2$ ensemble in order to gain insights into its microscopic origin.
Most importantly, the gralmonium’s resilience to magnetic fields offers a promising path forward in hybrid quantum architectures, facilitating seamless integration with magnetic-field-sensitive systems such as spins~\cite{Gunzler2025Feb}, magnons, or topological materials.

\section*{Data Availability}
All data supporting the findings of this study are available in~\cite{Gunzler2025Oct}.

\section*{Acknowledgements}
We are grateful to A. Shnirman and P. Bertet for fruitful discussions and we acknowledge technical support from S. Diewald, M.K. Gamer, and L. Radtke. 
Funding was provided by the German Research Foundation (DFG) via the Gottfried Wilhelm Leibniz-Award (ZVN-2020\_WE 4458-5) and by the European Research Council via project number 101118911 (DarkQuantum). 
N.Z. acknowledges funding from the Deutsche Forschungsgemeinschaft (DFG – German Research Foundation) under project number 450396347 (GeHoldeQED).
M.F. acknowledges funding from the European Union under the Horizon Europe Program, grant agreement number 101080152 (TruePA).
N.G. and M.S. acknowledge support from the German Ministry of Education and Research (BMBF) within the project GEQCOS (FKZ: 13N15683).
S.Ge. acknowledges support from the German Ministry of Education and Research (BMBF) within the project QSolid (FKZ: 13N16151).
Facilities use was supported by the Karlsruhe Nano Micro Facility (KNMFi) and KIT Nanostructure Service Laboratory (NSL). 
We acknowledge qKit for providing a convenient measurement software framework.


\bibliography{references}

\begin{thebibliography}{64}%
\makeatletter
\providecommand \@ifxundefined [1]{%
 \@ifx{#1\undefined}
}%
\providecommand \@ifnum [1]{%
 \ifnum #1\expandafter \@firstoftwo
 \else \expandafter \@secondoftwo
 \fi
}%
\providecommand \@ifx [1]{%
 \ifx #1\expandafter \@firstoftwo
 \else \expandafter \@secondoftwo
 \fi
}%
\providecommand \natexlab [1]{#1}%
\providecommand \enquote  [1]{``#1''}%
\providecommand \bibnamefont  [1]{#1}%
\providecommand \bibfnamefont [1]{#1}%
\providecommand \citenamefont [1]{#1}%
\providecommand \href@noop [0]{\@secondoftwo}%
\providecommand \href [0]{\begingroup \@sanitize@url \@href}%
\providecommand \@href[1]{\@@startlink{#1}\@@href}%
\providecommand \@@href[1]{\endgroup#1\@@endlink}%
\providecommand \@sanitize@url [0]{\catcode `\\12\catcode `\$12\catcode `\&12\catcode `\#12\catcode `\^12\catcode `\_12\catcode `\%12\relax}%
\providecommand \@@startlink[1]{}%
\providecommand \@@endlink[0]{}%
\providecommand \url  [0]{\begingroup\@sanitize@url \@url }%
\providecommand \@url [1]{\endgroup\@href {#1}{\urlprefix }}%
\providecommand \urlprefix  [0]{URL }%
\providecommand \Eprint [0]{\href }%
\providecommand \doibase [0]{https://doi.org/}%
\providecommand \selectlanguage [0]{\@gobble}%
\providecommand \bibinfo  [0]{\@secondoftwo}%
\providecommand \bibfield  [0]{\@secondoftwo}%
\providecommand \translation [1]{[#1]}%
\providecommand \BibitemOpen [0]{}%
\providecommand \bibitemStop [0]{}%
\providecommand \bibitemNoStop [0]{.\EOS\space}%
\providecommand \EOS [0]{\spacefactor3000\relax}%
\providecommand \BibitemShut  [1]{\csname bibitem#1\endcsname}%
\let\auto@bib@innerbib\@empty
\bibitem [{\citenamefont {Blais}\ \emph {et~al.}(2021)\citenamefont {Blais}, \citenamefont {Grimsmo}, \citenamefont {Girvin},\ and\ \citenamefont {Wallraff}}]{Blais2021May}%
  \BibitemOpen
  \bibfield  {author} {\bibinfo {author} {\bibfnamefont {A.}~\bibnamefont {Blais}}, \bibinfo {author} {\bibfnamefont {A.~L.}\ \bibnamefont {Grimsmo}}, \bibinfo {author} {\bibfnamefont {S.~M.}\ \bibnamefont {Girvin}},\ and\ \bibinfo {author} {\bibfnamefont {A.}~\bibnamefont {Wallraff}},\ }\bibfield  {title} {\bibinfo {title} {{Circuit quantum electrodynamics}},\ }\href {https://doi.org/10.1103/RevModPhys.93.025005} {\bibfield  {journal} {\bibinfo  {journal} {Rev. Mod. Phys.}\ }\textbf {\bibinfo {volume} {93}},\ \bibinfo {pages} {025005} (\bibinfo {year} {2021})}\BibitemShut {NoStop}%
\bibitem [{\citenamefont {Vijay}\ \emph {et~al.}(2011)\citenamefont {Vijay}, \citenamefont {Slichter},\ and\ \citenamefont {Siddiqi}}]{Vijay2011Mar}%
  \BibitemOpen
  \bibfield  {author} {\bibinfo {author} {\bibfnamefont {R.}~\bibnamefont {Vijay}}, \bibinfo {author} {\bibfnamefont {D.~H.}\ \bibnamefont {Slichter}},\ and\ \bibinfo {author} {\bibfnamefont {I.}~\bibnamefont {Siddiqi}},\ }\bibfield  {title} {\bibinfo {title} {{Observation of Quantum Jumps in a Superconducting Artificial Atom}},\ }\href {https://doi.org/10.1103/PhysRevLett.106.110502} {\bibfield  {journal} {\bibinfo  {journal} {Phys. Rev. Lett.}\ }\textbf {\bibinfo {volume} {106}},\ \bibinfo {pages} {110502} (\bibinfo {year} {2011})}\BibitemShut {NoStop}%
\bibitem [{\citenamefont {Ficheux}\ \emph {et~al.}(2018)\citenamefont {Ficheux}, \citenamefont {Jezouin}, \citenamefont {Leghtas},\ and\ \citenamefont {Huard}}]{Ficheux2018May}%
  \BibitemOpen
  \bibfield  {author} {\bibinfo {author} {\bibfnamefont {Q.}~\bibnamefont {Ficheux}}, \bibinfo {author} {\bibfnamefont {S.}~\bibnamefont {Jezouin}}, \bibinfo {author} {\bibfnamefont {Z.}~\bibnamefont {Leghtas}},\ and\ \bibinfo {author} {\bibfnamefont {B.}~\bibnamefont {Huard}},\ }\bibfield  {title} {\bibinfo {title} {{Dynamics of a qubit while simultaneously monitoring its relaxation and dephasing}},\ }\href {https://doi.org/10.1038/s41467-018-04372-9} {\bibfield  {journal} {\bibinfo  {journal} {Nat. Commun.}\ }\textbf {\bibinfo {volume} {9}},\ \bibinfo {pages} {1} (\bibinfo {year} {2018})}\BibitemShut {NoStop}%
\bibitem [{\citenamefont {Minev}\ \emph {et~al.}(2019)\citenamefont {Minev}, \citenamefont {Mundhada}, \citenamefont {Shankar}, \citenamefont {Reinhold}, \citenamefont {Guti{\ifmmode\acute{e}\else\'{e}\fi}rrez-J{\ifmmode\acute{a}\else\'{a}\fi}uregui}, \citenamefont {Schoelkopf}, \citenamefont {Mirrahimi}, \citenamefont {Carmichael},\ and\ \citenamefont {Devoret}}]{Minev2019Jun}%
  \BibitemOpen
  \bibfield  {author} {\bibinfo {author} {\bibfnamefont {Z.~K.}\ \bibnamefont {Minev}}, \bibinfo {author} {\bibfnamefont {S.~O.}\ \bibnamefont {Mundhada}}, \bibinfo {author} {\bibfnamefont {S.}~\bibnamefont {Shankar}}, \bibinfo {author} {\bibfnamefont {P.}~\bibnamefont {Reinhold}}, \bibinfo {author} {\bibfnamefont {R.}~\bibnamefont {Guti{\ifmmode\acute{e}\else\'{e}\fi}rrez-J{\ifmmode\acute{a}\else\'{a}\fi}uregui}}, \bibinfo {author} {\bibfnamefont {R.~J.}\ \bibnamefont {Schoelkopf}}, \bibinfo {author} {\bibfnamefont {M.}~\bibnamefont {Mirrahimi}}, \bibinfo {author} {\bibfnamefont {H.~J.}\ \bibnamefont {Carmichael}},\ and\ \bibinfo {author} {\bibfnamefont {M.~H.}\ \bibnamefont {Devoret}},\ }\bibfield  {title} {\bibinfo {title} {{To catch and reverse a quantum jump mid-flight}},\ }\href {https://doi.org/10.1038/s41586-019-1287-z} {\bibfield  {journal} {\bibinfo  {journal} {Nature}\ }\textbf {\bibinfo {volume} {570}},\ \bibinfo {pages} {200} (\bibinfo {year} {2019})}\BibitemShut {NoStop}%
\bibitem [{\citenamefont {Vijay}\ \emph {et~al.}(2012)\citenamefont {Vijay}, \citenamefont {Macklin}, \citenamefont {Slichter}, \citenamefont {Weber}, \citenamefont {Murch}, \citenamefont {Naik}, \citenamefont {Korotkov},\ and\ \citenamefont {Siddiqi}}]{Vijay2012Oct}%
  \BibitemOpen
  \bibfield  {author} {\bibinfo {author} {\bibfnamefont {R.}~\bibnamefont {Vijay}}, \bibinfo {author} {\bibfnamefont {C.}~\bibnamefont {Macklin}}, \bibinfo {author} {\bibfnamefont {D.~H.}\ \bibnamefont {Slichter}}, \bibinfo {author} {\bibfnamefont {S.~J.}\ \bibnamefont {Weber}}, \bibinfo {author} {\bibfnamefont {K.~W.}\ \bibnamefont {Murch}}, \bibinfo {author} {\bibfnamefont {R.}~\bibnamefont {Naik}}, \bibinfo {author} {\bibfnamefont {A.~N.}\ \bibnamefont {Korotkov}},\ and\ \bibinfo {author} {\bibfnamefont {I.}~\bibnamefont {Siddiqi}},\ }\bibfield  {title} {\bibinfo {title} {{Stabilizing Rabi oscillations in a superconducting qubit using quantum feedback}},\ }\href {https://doi.org/10.1038/nature11505} {\bibfield  {journal} {\bibinfo  {journal} {Nature}\ }\textbf {\bibinfo {volume} {490}},\ \bibinfo {pages} {77} (\bibinfo {year} {2012})}\BibitemShut {NoStop}%
\bibitem [{\citenamefont {Sivak}\ \emph {et~al.}(2023)\citenamefont {Sivak}, \citenamefont {Eickbusch}, \citenamefont {Royer}, \citenamefont {Singh}, \citenamefont {Tsioutsios}, \citenamefont {Ganjam}, \citenamefont {Miano}, \citenamefont {Brock}, \citenamefont {Ding}, \citenamefont {Frunzio}, \citenamefont {Girvin}, \citenamefont {Schoelkopf},\ and\ \citenamefont {Devoret}}]{Sivak2023Apr}%
  \BibitemOpen
  \bibfield  {author} {\bibinfo {author} {\bibfnamefont {V.~V.}\ \bibnamefont {Sivak}}, \bibinfo {author} {\bibfnamefont {A.}~\bibnamefont {Eickbusch}}, \bibinfo {author} {\bibfnamefont {B.}~\bibnamefont {Royer}}, \bibinfo {author} {\bibfnamefont {S.}~\bibnamefont {Singh}}, \bibinfo {author} {\bibfnamefont {I.}~\bibnamefont {Tsioutsios}}, \bibinfo {author} {\bibfnamefont {S.}~\bibnamefont {Ganjam}}, \bibinfo {author} {\bibfnamefont {A.}~\bibnamefont {Miano}}, \bibinfo {author} {\bibfnamefont {B.~L.}\ \bibnamefont {Brock}}, \bibinfo {author} {\bibfnamefont {A.~Z.}\ \bibnamefont {Ding}}, \bibinfo {author} {\bibfnamefont {L.}~\bibnamefont {Frunzio}}, \bibinfo {author} {\bibfnamefont {S.~M.}\ \bibnamefont {Girvin}}, \bibinfo {author} {\bibfnamefont {R.~J.}\ \bibnamefont {Schoelkopf}},\ and\ \bibinfo {author} {\bibfnamefont {M.~H.}\ \bibnamefont {Devoret}},\ }\bibfield  {title} {\bibinfo {title} {{Real-time quantum error correction beyond break-even}},\ }\href {https://doi.org/10.1038/s41586-023-05782-6}
  {\bibfield  {journal} {\bibinfo  {journal} {Nature}\ }\textbf {\bibinfo {volume} {616}},\ \bibinfo {pages} {50} (\bibinfo {year} {2023})}\BibitemShut {NoStop}%
\bibitem [{\citenamefont {Krinner}\ \emph {et~al.}(2022)\citenamefont {Krinner}, \citenamefont {Lacroix}, \citenamefont {Remm}, \citenamefont {Di~Paolo}, \citenamefont {Genois}, \citenamefont {Leroux}, \citenamefont {Hellings}, \citenamefont {Lazar}, \citenamefont {Swiadek}, \citenamefont {Herrmann}, \citenamefont {Norris}, \citenamefont {Andersen}, \citenamefont {M{\ifmmode\ddot{u}\else\"{u}\fi}ller}, \citenamefont {Blais}, \citenamefont {Eichler},\ and\ \citenamefont {Wallraff}}]{Krinner2022May}%
  \BibitemOpen
  \bibfield  {author} {\bibinfo {author} {\bibfnamefont {S.}~\bibnamefont {Krinner}}, \bibinfo {author} {\bibfnamefont {N.}~\bibnamefont {Lacroix}}, \bibinfo {author} {\bibfnamefont {A.}~\bibnamefont {Remm}}, \bibinfo {author} {\bibfnamefont {A.}~\bibnamefont {Di~Paolo}}, \bibinfo {author} {\bibfnamefont {E.}~\bibnamefont {Genois}}, \bibinfo {author} {\bibfnamefont {C.}~\bibnamefont {Leroux}}, \bibinfo {author} {\bibfnamefont {C.}~\bibnamefont {Hellings}}, \bibinfo {author} {\bibfnamefont {S.}~\bibnamefont {Lazar}}, \bibinfo {author} {\bibfnamefont {F.}~\bibnamefont {Swiadek}}, \bibinfo {author} {\bibfnamefont {J.}~\bibnamefont {Herrmann}}, \bibinfo {author} {\bibfnamefont {G.~J.}\ \bibnamefont {Norris}}, \bibinfo {author} {\bibfnamefont {C.~K.}\ \bibnamefont {Andersen}}, \bibinfo {author} {\bibfnamefont {M.}~\bibnamefont {M{\ifmmode\ddot{u}\else\"{u}\fi}ller}}, \bibinfo {author} {\bibfnamefont {A.}~\bibnamefont {Blais}}, \bibinfo {author} {\bibfnamefont {C.}~\bibnamefont {Eichler}},\ and\ \bibinfo {author}
  {\bibfnamefont {A.}~\bibnamefont {Wallraff}},\ }\bibfield  {title} {\bibinfo {title} {{Realizing repeated quantum error correction in a distance-three surface code}},\ }\href {https://doi.org/10.1038/s41586-022-04566-8} {\bibfield  {journal} {\bibinfo  {journal} {Nature}\ }\textbf {\bibinfo {volume} {605}},\ \bibinfo {pages} {669} (\bibinfo {year} {2022})}\BibitemShut {NoStop}%
\bibitem [{\citenamefont {Saxberg}\ \emph {et~al.}(2022)\citenamefont {Saxberg}, \citenamefont {Vrajitoarea}, \citenamefont {Roberts}, \citenamefont {Panetta}, \citenamefont {Simon},\ and\ \citenamefont {Schuster}}]{Saxberg2022Dec}%
  \BibitemOpen
  \bibfield  {author} {\bibinfo {author} {\bibfnamefont {B.}~\bibnamefont {Saxberg}}, \bibinfo {author} {\bibfnamefont {A.}~\bibnamefont {Vrajitoarea}}, \bibinfo {author} {\bibfnamefont {G.}~\bibnamefont {Roberts}}, \bibinfo {author} {\bibfnamefont {M.~G.}\ \bibnamefont {Panetta}}, \bibinfo {author} {\bibfnamefont {J.}~\bibnamefont {Simon}},\ and\ \bibinfo {author} {\bibfnamefont {D.~I.}\ \bibnamefont {Schuster}},\ }\bibfield  {title} {\bibinfo {title} {{Disorder-assisted assembly of strongly correlated fluids of light}},\ }\href {https://doi.org/10.1038/s41586-022-05357-x} {\bibfield  {journal} {\bibinfo  {journal} {Nature}\ }\textbf {\bibinfo {volume} {612}},\ \bibinfo {pages} {435} (\bibinfo {year} {2022})}\BibitemShut {NoStop}%
\bibitem [{\citenamefont {Bild}\ \emph {et~al.}(2023)\citenamefont {Bild}, \citenamefont {Fadel}, \citenamefont {Yang}, \citenamefont {von L{\ifmmode\ddot{u}\else\"{u}\fi}pke}, \citenamefont {Martin}, \citenamefont {Bruno},\ and\ \citenamefont {Chu}}]{Bild2023Apr}%
  \BibitemOpen
  \bibfield  {author} {\bibinfo {author} {\bibfnamefont {M.}~\bibnamefont {Bild}}, \bibinfo {author} {\bibfnamefont {M.}~\bibnamefont {Fadel}}, \bibinfo {author} {\bibfnamefont {Y.}~\bibnamefont {Yang}}, \bibinfo {author} {\bibfnamefont {U.}~\bibnamefont {von L{\ifmmode\ddot{u}\else\"{u}\fi}pke}}, \bibinfo {author} {\bibfnamefont {P.}~\bibnamefont {Martin}}, \bibinfo {author} {\bibfnamefont {A.}~\bibnamefont {Bruno}},\ and\ \bibinfo {author} {\bibfnamefont {Y.}~\bibnamefont {Chu}},\ }\bibfield  {title} {\bibinfo {title} {{Schr{\ifmmode\ddot{o}\else\"{o}\fi}dinger cat states of a 16-microgram mechanical oscillator}},\ }\href {https://doi.org/10.1126/science.adf7553} {\bibfield  {journal} {\bibinfo  {journal} {Science}\ }\textbf {\bibinfo {volume} {380}},\ \bibinfo {pages} {274} (\bibinfo {year} {2023})}\BibitemShut {NoStop}%
\bibitem [{\citenamefont {Storz}\ \emph {et~al.}(2023)\citenamefont {Storz}, \citenamefont {Sch{\ifmmode\ddot{a}\else\"{a}\fi}r}, \citenamefont {Kulikov}, \citenamefont {Magnard}, \citenamefont {Kurpiers}, \citenamefont {L{\ifmmode\ddot{u}\else\"{u}\fi}tolf}, \citenamefont {Walter}, \citenamefont {Copetudo}, \citenamefont {Reuer}, \citenamefont {Akin}, \citenamefont {Besse}, \citenamefont {Gabureac}, \citenamefont {Norris}, \citenamefont {Rosario}, \citenamefont {Martin}, \citenamefont {Martinez}, \citenamefont {Amaya}, \citenamefont {Mitchell}, \citenamefont {Abellan}, \citenamefont {Bancal}, \citenamefont {Sangouard}, \citenamefont {Royer}, \citenamefont {Blais},\ and\ \citenamefont {Wallraff}}]{Storz2023May}%
  \BibitemOpen
  \bibfield  {author} {\bibinfo {author} {\bibfnamefont {S.}~\bibnamefont {Storz}}, \bibinfo {author} {\bibfnamefont {J.}~\bibnamefont {Sch{\ifmmode\ddot{a}\else\"{a}\fi}r}}, \bibinfo {author} {\bibfnamefont {A.}~\bibnamefont {Kulikov}}, \bibinfo {author} {\bibfnamefont {P.}~\bibnamefont {Magnard}}, \bibinfo {author} {\bibfnamefont {P.}~\bibnamefont {Kurpiers}}, \bibinfo {author} {\bibfnamefont {J.}~\bibnamefont {L{\ifmmode\ddot{u}\else\"{u}\fi}tolf}}, \bibinfo {author} {\bibfnamefont {T.}~\bibnamefont {Walter}}, \bibinfo {author} {\bibfnamefont {A.}~\bibnamefont {Copetudo}}, \bibinfo {author} {\bibfnamefont {K.}~\bibnamefont {Reuer}}, \bibinfo {author} {\bibfnamefont {A.}~\bibnamefont {Akin}}, \bibinfo {author} {\bibfnamefont {J.-C.}\ \bibnamefont {Besse}}, \bibinfo {author} {\bibfnamefont {M.}~\bibnamefont {Gabureac}}, \bibinfo {author} {\bibfnamefont {G.~J.}\ \bibnamefont {Norris}}, \bibinfo {author} {\bibfnamefont {A.}~\bibnamefont {Rosario}}, \bibinfo {author} {\bibfnamefont {F.}~\bibnamefont {Martin}},
  \bibinfo {author} {\bibfnamefont {J.}~\bibnamefont {Martinez}}, \bibinfo {author} {\bibfnamefont {W.}~\bibnamefont {Amaya}}, \bibinfo {author} {\bibfnamefont {M.~W.}\ \bibnamefont {Mitchell}}, \bibinfo {author} {\bibfnamefont {C.}~\bibnamefont {Abellan}}, \bibinfo {author} {\bibfnamefont {J.-D.}\ \bibnamefont {Bancal}}, \bibinfo {author} {\bibfnamefont {N.}~\bibnamefont {Sangouard}}, \bibinfo {author} {\bibfnamefont {B.}~\bibnamefont {Royer}}, \bibinfo {author} {\bibfnamefont {A.}~\bibnamefont {Blais}},\ and\ \bibinfo {author} {\bibfnamefont {A.}~\bibnamefont {Wallraff}},\ }\bibfield  {title} {\bibinfo {title} {{Loophole-free Bell inequality violation with superconducting circuits}},\ }\href {https://doi.org/10.1038/s41586-023-05885-0} {\bibfield  {journal} {\bibinfo  {journal} {Nature}\ }\textbf {\bibinfo {volume} {617}},\ \bibinfo {pages} {265} (\bibinfo {year} {2023})}\BibitemShut {NoStop}%
\bibitem [{\citenamefont {Landig}\ \emph {et~al.}(2018)\citenamefont {Landig}, \citenamefont {Koski}, \citenamefont {Scarlino}, \citenamefont {Mendes}, \citenamefont {Blais}, \citenamefont {Reichl}, \citenamefont {Wegscheider}, \citenamefont {Wallraff}, \citenamefont {Ensslin},\ and\ \citenamefont {Ihn}}]{Landig2018Aug}%
  \BibitemOpen
  \bibfield  {author} {\bibinfo {author} {\bibfnamefont {A.~J.}\ \bibnamefont {Landig}}, \bibinfo {author} {\bibfnamefont {J.~V.}\ \bibnamefont {Koski}}, \bibinfo {author} {\bibfnamefont {P.}~\bibnamefont {Scarlino}}, \bibinfo {author} {\bibfnamefont {U.~C.}\ \bibnamefont {Mendes}}, \bibinfo {author} {\bibfnamefont {A.}~\bibnamefont {Blais}}, \bibinfo {author} {\bibfnamefont {C.}~\bibnamefont {Reichl}}, \bibinfo {author} {\bibfnamefont {W.}~\bibnamefont {Wegscheider}}, \bibinfo {author} {\bibfnamefont {A.}~\bibnamefont {Wallraff}}, \bibinfo {author} {\bibfnamefont {K.}~\bibnamefont {Ensslin}},\ and\ \bibinfo {author} {\bibfnamefont {T.}~\bibnamefont {Ihn}},\ }\bibfield  {title} {\bibinfo {title} {{Coherent spin{\textendash}photon coupling using a resonant exchange qubit}},\ }\href {https://doi.org/10.1038/s41586-018-0365-y} {\bibfield  {journal} {\bibinfo  {journal} {Nature}\ }\textbf {\bibinfo {volume} {560}},\ \bibinfo {pages} {179} (\bibinfo {year} {2018})}\BibitemShut {NoStop}%
\bibitem [{\citenamefont {Samkharadze}\ \emph {et~al.}(2018)\citenamefont {Samkharadze}, \citenamefont {Zheng}, \citenamefont {Kalhor}, \citenamefont {Brousse}, \citenamefont {Sammak}, \citenamefont {Mendes}, \citenamefont {Blais}, \citenamefont {Scappucci},\ and\ \citenamefont {Vandersypen}}]{Samkharadze2018Jan}%
  \BibitemOpen
  \bibfield  {author} {\bibinfo {author} {\bibfnamefont {N.}~\bibnamefont {Samkharadze}}, \bibinfo {author} {\bibfnamefont {G.}~\bibnamefont {Zheng}}, \bibinfo {author} {\bibfnamefont {N.}~\bibnamefont {Kalhor}}, \bibinfo {author} {\bibfnamefont {D.}~\bibnamefont {Brousse}}, \bibinfo {author} {\bibfnamefont {A.}~\bibnamefont {Sammak}}, \bibinfo {author} {\bibfnamefont {U.~C.}\ \bibnamefont {Mendes}}, \bibinfo {author} {\bibfnamefont {A.}~\bibnamefont {Blais}}, \bibinfo {author} {\bibfnamefont {G.}~\bibnamefont {Scappucci}},\ and\ \bibinfo {author} {\bibfnamefont {L.~M.~K.}\ \bibnamefont {Vandersypen}},\ }\bibfield  {title} {\bibinfo {title} {{Strong spin-photon coupling in silicon}},\ }\href {https://doi.org/10.1126/science.aar4054} {\bibfield  {journal} {\bibinfo  {journal} {Science}\ }\textbf {\bibinfo {volume} {359}},\ \bibinfo {pages} {1123} (\bibinfo {year} {2018})}\BibitemShut {NoStop}%
\bibitem [{\citenamefont {Mi}\ \emph {et~al.}(2018)\citenamefont {Mi}, \citenamefont {Benito}, \citenamefont {Putz}, \citenamefont {Zajac}, \citenamefont {Taylor}, \citenamefont {Burkard},\ and\ \citenamefont {Petta}}]{Mi2018Mar}%
  \BibitemOpen
  \bibfield  {author} {\bibinfo {author} {\bibfnamefont {X.}~\bibnamefont {Mi}}, \bibinfo {author} {\bibfnamefont {M.}~\bibnamefont {Benito}}, \bibinfo {author} {\bibfnamefont {S.}~\bibnamefont {Putz}}, \bibinfo {author} {\bibfnamefont {D.~M.}\ \bibnamefont {Zajac}}, \bibinfo {author} {\bibfnamefont {J.~M.}\ \bibnamefont {Taylor}}, \bibinfo {author} {\bibfnamefont {G.}~\bibnamefont {Burkard}},\ and\ \bibinfo {author} {\bibfnamefont {J.~R.}\ \bibnamefont {Petta}},\ }\bibfield  {title} {\bibinfo {title} {{A coherent spin{\textendash}photon interface in silicon}},\ }\href {https://doi.org/10.1038/nature25769} {\bibfield  {journal} {\bibinfo  {journal} {Nature}\ }\textbf {\bibinfo {volume} {555}},\ \bibinfo {pages} {599} (\bibinfo {year} {2018})}\BibitemShut {NoStop}%
\bibitem [{\citenamefont {Bienfait}\ \emph {et~al.}(2016{\natexlab{a}})\citenamefont {Bienfait}, \citenamefont {Pla}, \citenamefont {Kubo}, \citenamefont {Zhou}, \citenamefont {Stern}, \citenamefont {Lo}, \citenamefont {Weis}, \citenamefont {Schenkel}, \citenamefont {Vion}, \citenamefont {Esteve}, \citenamefont {Morton},\ and\ \citenamefont {Bertet}}]{Bienfait2016Mar}%
  \BibitemOpen
  \bibfield  {author} {\bibinfo {author} {\bibfnamefont {A.}~\bibnamefont {Bienfait}}, \bibinfo {author} {\bibfnamefont {J.~J.}\ \bibnamefont {Pla}}, \bibinfo {author} {\bibfnamefont {Y.}~\bibnamefont {Kubo}}, \bibinfo {author} {\bibfnamefont {X.}~\bibnamefont {Zhou}}, \bibinfo {author} {\bibfnamefont {M.}~\bibnamefont {Stern}}, \bibinfo {author} {\bibfnamefont {C.~C.}\ \bibnamefont {Lo}}, \bibinfo {author} {\bibfnamefont {C.~D.}\ \bibnamefont {Weis}}, \bibinfo {author} {\bibfnamefont {T.}~\bibnamefont {Schenkel}}, \bibinfo {author} {\bibfnamefont {D.}~\bibnamefont {Vion}}, \bibinfo {author} {\bibfnamefont {D.}~\bibnamefont {Esteve}}, \bibinfo {author} {\bibfnamefont {J.~J.~L.}\ \bibnamefont {Morton}},\ and\ \bibinfo {author} {\bibfnamefont {P.}~\bibnamefont {Bertet}},\ }\bibfield  {title} {\bibinfo {title} {{Controlling spin relaxation with a cavity}},\ }\href {https://doi.org/10.1038/nature16944} {\bibfield  {journal} {\bibinfo  {journal} {Nature}\ }\textbf {\bibinfo {volume} {531}},\ \bibinfo {pages} {74}
  (\bibinfo {year} {2016}{\natexlab{a}})}\BibitemShut {NoStop}%
\bibitem [{\citenamefont {Eichler}\ \emph {et~al.}(2017)\citenamefont {Eichler}, \citenamefont {Sigillito}, \citenamefont {Lyon},\ and\ \citenamefont {Petta}}]{Eichler2017Jan}%
  \BibitemOpen
  \bibfield  {author} {\bibinfo {author} {\bibfnamefont {C.}~\bibnamefont {Eichler}}, \bibinfo {author} {\bibfnamefont {A.~J.}\ \bibnamefont {Sigillito}}, \bibinfo {author} {\bibfnamefont {S.~A.}\ \bibnamefont {Lyon}},\ and\ \bibinfo {author} {\bibfnamefont {J.~R.}\ \bibnamefont {Petta}},\ }\bibfield  {title} {\bibinfo {title} {{Electron Spin Resonance at the Level of $1{0}^{4}$ Spins Using Low Impedance Superconducting Resonators}},\ }\href {https://doi.org/10.1103/PhysRevLett.118.037701} {\bibfield  {journal} {\bibinfo  {journal} {Phys. Rev. Lett.}\ }\textbf {\bibinfo {volume} {118}},\ \bibinfo {pages} {037701} (\bibinfo {year} {2017})}\BibitemShut {NoStop}%
\bibitem [{\citenamefont {Bienfait}\ \emph {et~al.}(2016{\natexlab{b}})\citenamefont {Bienfait}, \citenamefont {Pla}, \citenamefont {Kubo}, \citenamefont {Stern}, \citenamefont {Zhou}, \citenamefont {Lo}, \citenamefont {Weis}, \citenamefont {Schenkel}, \citenamefont {Thewalt}, \citenamefont {Vion}, \citenamefont {Esteve}, \citenamefont {Julsgaard}, \citenamefont {M{\o}lmer}, \citenamefont {Morton},\ and\ \citenamefont {Bertet}}]{Bienfait2016_2}%
  \BibitemOpen
  \bibfield  {author} {\bibinfo {author} {\bibfnamefont {A.}~\bibnamefont {Bienfait}}, \bibinfo {author} {\bibfnamefont {J.~J.}\ \bibnamefont {Pla}}, \bibinfo {author} {\bibfnamefont {Y.}~\bibnamefont {Kubo}}, \bibinfo {author} {\bibfnamefont {M.}~\bibnamefont {Stern}}, \bibinfo {author} {\bibfnamefont {X.}~\bibnamefont {Zhou}}, \bibinfo {author} {\bibfnamefont {C.~C.}\ \bibnamefont {Lo}}, \bibinfo {author} {\bibfnamefont {C.~D.}\ \bibnamefont {Weis}}, \bibinfo {author} {\bibfnamefont {T.}~\bibnamefont {Schenkel}}, \bibinfo {author} {\bibfnamefont {M.~L.~W.}\ \bibnamefont {Thewalt}}, \bibinfo {author} {\bibfnamefont {D.}~\bibnamefont {Vion}}, \bibinfo {author} {\bibfnamefont {D.}~\bibnamefont {Esteve}}, \bibinfo {author} {\bibfnamefont {B.}~\bibnamefont {Julsgaard}}, \bibinfo {author} {\bibfnamefont {K.}~\bibnamefont {M{\o}lmer}}, \bibinfo {author} {\bibfnamefont {J.~J.~L.}\ \bibnamefont {Morton}},\ and\ \bibinfo {author} {\bibfnamefont {P.}~\bibnamefont {Bertet}},\ }\bibfield  {title} {\bibinfo {title}
  {{Reaching the quantum limit of sensitivity in electron spin resonance}},\ }\href {https://doi.org/10.1038/nnano.2015.282} {\bibfield  {journal} {\bibinfo  {journal} {Nat. Nanotechnol.}\ }\textbf {\bibinfo {volume} {11}},\ \bibinfo {pages} {253} (\bibinfo {year} {2016}{\natexlab{b}})}\BibitemShut {NoStop}%
\bibitem [{\citenamefont {Wang}\ \emph {et~al.}(2023)\citenamefont {Wang}, \citenamefont {Balembois}, \citenamefont {Ran{\ifmmode\check{c}\else\v{c}\fi}i{\ifmmode\acute{c}\else\'{c}\fi}}, \citenamefont {Billaud}, \citenamefont {Le~Dantec}, \citenamefont {Ferrier}, \citenamefont {Goldner}, \citenamefont {Bertaina}, \citenamefont {Chaneli{\ifmmode\grave{e}\else\`{e}\fi}re}, \citenamefont {Esteve}, \citenamefont {Vion}, \citenamefont {Bertet},\ and\ \citenamefont {Flurin}}]{Wang2023Jul}%
  \BibitemOpen
  \bibfield  {author} {\bibinfo {author} {\bibfnamefont {Z.}~\bibnamefont {Wang}}, \bibinfo {author} {\bibfnamefont {L.}~\bibnamefont {Balembois}}, \bibinfo {author} {\bibfnamefont {M.}~\bibnamefont {Ran{\ifmmode\check{c}\else\v{c}\fi}i{\ifmmode\acute{c}\else\'{c}\fi}}}, \bibinfo {author} {\bibfnamefont {E.}~\bibnamefont {Billaud}}, \bibinfo {author} {\bibfnamefont {M.}~\bibnamefont {Le~Dantec}}, \bibinfo {author} {\bibfnamefont {A.}~\bibnamefont {Ferrier}}, \bibinfo {author} {\bibfnamefont {P.}~\bibnamefont {Goldner}}, \bibinfo {author} {\bibfnamefont {S.}~\bibnamefont {Bertaina}}, \bibinfo {author} {\bibfnamefont {T.}~\bibnamefont {Chaneli{\ifmmode\grave{e}\else\`{e}\fi}re}}, \bibinfo {author} {\bibfnamefont {D.}~\bibnamefont {Esteve}}, \bibinfo {author} {\bibfnamefont {D.}~\bibnamefont {Vion}}, \bibinfo {author} {\bibfnamefont {P.}~\bibnamefont {Bertet}},\ and\ \bibinfo {author} {\bibfnamefont {E.}~\bibnamefont {Flurin}},\ }\bibfield  {title} {\bibinfo {title} {{Single-electron spin resonance detection by
  microwave photon counting}},\ }\href {https://doi.org/10.1038/s41586-023-06097-2} {\bibfield  {journal} {\bibinfo  {journal} {Nature}\ }\textbf {\bibinfo {volume} {619}},\ \bibinfo {pages} {276} (\bibinfo {year} {2023})}\BibitemShut {NoStop}%
\bibitem [{\citenamefont {O'Sullivan}\ \emph {et~al.}(2024)\citenamefont {O'Sullivan}, \citenamefont {Travesedo}, \citenamefont {Pallegoix}, \citenamefont {Huang}, \citenamefont {May}, \citenamefont {Yavkin}, \citenamefont {Hogan}, \citenamefont {Lin}, \citenamefont {Liu}, \citenamefont {Chaneliere}, \citenamefont {Bertaina}, \citenamefont {Goldner}, \citenamefont {Esteve}, \citenamefont {Vion}, \citenamefont {Abgrall}, \citenamefont {Bertet},\ and\ \citenamefont {Flurin}}]{OSullivan2024Oct}%
  \BibitemOpen
  \bibfield  {author} {\bibinfo {author} {\bibfnamefont {J.}~\bibnamefont {O'Sullivan}}, \bibinfo {author} {\bibfnamefont {J.}~\bibnamefont {Travesedo}}, \bibinfo {author} {\bibfnamefont {L.}~\bibnamefont {Pallegoix}}, \bibinfo {author} {\bibfnamefont {Z.~W.}\ \bibnamefont {Huang}}, \bibinfo {author} {\bibfnamefont {A.}~\bibnamefont {May}}, \bibinfo {author} {\bibfnamefont {B.}~\bibnamefont {Yavkin}}, \bibinfo {author} {\bibfnamefont {P.}~\bibnamefont {Hogan}}, \bibinfo {author} {\bibfnamefont {S.}~\bibnamefont {Lin}}, \bibinfo {author} {\bibfnamefont {R.}~\bibnamefont {Liu}}, \bibinfo {author} {\bibfnamefont {T.}~\bibnamefont {Chaneliere}}, \bibinfo {author} {\bibfnamefont {S.}~\bibnamefont {Bertaina}}, \bibinfo {author} {\bibfnamefont {P.}~\bibnamefont {Goldner}}, \bibinfo {author} {\bibfnamefont {D.}~\bibnamefont {Esteve}}, \bibinfo {author} {\bibfnamefont {D.}~\bibnamefont {Vion}}, \bibinfo {author} {\bibfnamefont {P.}~\bibnamefont {Abgrall}}, \bibinfo {author} {\bibfnamefont {P.}~\bibnamefont {Bertet}},\
  and\ \bibinfo {author} {\bibfnamefont {E.}~\bibnamefont {Flurin}},\ }\bibfield  {title} {\bibinfo {title} {{Individual solid-state nuclear spin qubits with coherence exceeding seconds}},\ }\bibfield  {journal} {\bibinfo  {journal} {arXiv}\ }\href {https://doi.org/10.48550/arXiv.2410.10432} {10.48550/arXiv.2410.10432} (\bibinfo {year} {2024}),\ \Eprint {https://arxiv.org/abs/2410.10432} {2410.10432} \BibitemShut {NoStop}%
\bibitem [{\citenamefont {Lachance-Quirion}\ \emph {et~al.}(2020)\citenamefont {Lachance-Quirion}, \citenamefont {Wolski}, \citenamefont {Tabuchi}, \citenamefont {Kono}, \citenamefont {Usami},\ and\ \citenamefont {Nakamura}}]{Lachance-Quirion2020Jan}%
  \BibitemOpen
  \bibfield  {author} {\bibinfo {author} {\bibfnamefont {D.}~\bibnamefont {Lachance-Quirion}}, \bibinfo {author} {\bibfnamefont {S.~P.}\ \bibnamefont {Wolski}}, \bibinfo {author} {\bibfnamefont {Y.}~\bibnamefont {Tabuchi}}, \bibinfo {author} {\bibfnamefont {S.}~\bibnamefont {Kono}}, \bibinfo {author} {\bibfnamefont {K.}~\bibnamefont {Usami}},\ and\ \bibinfo {author} {\bibfnamefont {Y.}~\bibnamefont {Nakamura}},\ }\bibfield  {title} {\bibinfo {title} {{Entanglement-based single-shot detection of a single magnon with a superconducting qubit}},\ }\href {https://doi.org/10.1126/science.aaz9236} {\bibfield  {journal} {\bibinfo  {journal} {Science}\ }\textbf {\bibinfo {volume} {367}},\ \bibinfo {pages} {425} (\bibinfo {year} {2020})}\BibitemShut {NoStop}%
\bibitem [{\citenamefont {Serniak}\ \emph {et~al.}(2019)\citenamefont {Serniak}, \citenamefont {Diamond}, \citenamefont {Hays}, \citenamefont {Fatemi}, \citenamefont {Shankar}, \citenamefont {Frunzio}, \citenamefont {Schoelkopf},\ and\ \citenamefont {Devoret}}]{Serniak2019Jul}%
  \BibitemOpen
  \bibfield  {author} {\bibinfo {author} {\bibfnamefont {K.}~\bibnamefont {Serniak}}, \bibinfo {author} {\bibfnamefont {S.}~\bibnamefont {Diamond}}, \bibinfo {author} {\bibfnamefont {M.}~\bibnamefont {Hays}}, \bibinfo {author} {\bibfnamefont {V.}~\bibnamefont {Fatemi}}, \bibinfo {author} {\bibfnamefont {S.}~\bibnamefont {Shankar}}, \bibinfo {author} {\bibfnamefont {L.}~\bibnamefont {Frunzio}}, \bibinfo {author} {\bibfnamefont {R.~J.}\ \bibnamefont {Schoelkopf}},\ and\ \bibinfo {author} {\bibfnamefont {M.~H.}\ \bibnamefont {Devoret}},\ }\bibfield  {title} {\bibinfo {title} {{Direct Dispersive Monitoring of Charge Parity in Offset-Charge-Sensitive Transmons}},\ }\href {https://doi.org/10.1103/PhysRevApplied.12.014052} {\bibfield  {journal} {\bibinfo  {journal} {Phys. Rev. Appl.}\ }\textbf {\bibinfo {volume} {12}},\ \bibinfo {pages} {014052} (\bibinfo {year} {2019})}\BibitemShut {NoStop}%
\bibitem [{\citenamefont {McEwen}\ \emph {et~al.}(2024)\citenamefont {McEwen}, \citenamefont {Miao}, \citenamefont {Atalaya}, \citenamefont {Bilmes}, \citenamefont {Crook}, \citenamefont {Bovaird}, \citenamefont {Kreikebaum}, \citenamefont {Zobrist}, \citenamefont {Jeffrey}, \citenamefont {Ying}, \citenamefont {Bengtsson}, \citenamefont {Chang}, \citenamefont {Dunsworth}, \citenamefont {Kelly}, \citenamefont {Zhang}, \citenamefont {Forati}, \citenamefont {Acharya}, \citenamefont {Iveland}, \citenamefont {Liu}, \citenamefont {Kim}, \citenamefont {Burkett}, \citenamefont {Megrant}, \citenamefont {Chen}, \citenamefont {Neill}, \citenamefont {Sank}, \citenamefont {Devoret},\ and\ \citenamefont {Opremcak}}]{McEwen2024Feb}%
  \BibitemOpen
  \bibfield  {author} {\bibinfo {author} {\bibfnamefont {M.}~\bibnamefont {McEwen}}, \bibinfo {author} {\bibfnamefont {K.~C.}\ \bibnamefont {Miao}}, \bibinfo {author} {\bibfnamefont {J.}~\bibnamefont {Atalaya}}, \bibinfo {author} {\bibfnamefont {A.}~\bibnamefont {Bilmes}}, \bibinfo {author} {\bibfnamefont {A.}~\bibnamefont {Crook}}, \bibinfo {author} {\bibfnamefont {J.}~\bibnamefont {Bovaird}}, \bibinfo {author} {\bibfnamefont {J.~M.}\ \bibnamefont {Kreikebaum}}, \bibinfo {author} {\bibfnamefont {N.}~\bibnamefont {Zobrist}}, \bibinfo {author} {\bibfnamefont {E.}~\bibnamefont {Jeffrey}}, \bibinfo {author} {\bibfnamefont {B.}~\bibnamefont {Ying}}, \bibinfo {author} {\bibfnamefont {A.}~\bibnamefont {Bengtsson}}, \bibinfo {author} {\bibfnamefont {H.-S.}\ \bibnamefont {Chang}}, \bibinfo {author} {\bibfnamefont {A.}~\bibnamefont {Dunsworth}}, \bibinfo {author} {\bibfnamefont {J.}~\bibnamefont {Kelly}}, \bibinfo {author} {\bibfnamefont {Y.}~\bibnamefont {Zhang}}, \bibinfo {author} {\bibfnamefont {E.}~\bibnamefont
  {Forati}}, \bibinfo {author} {\bibfnamefont {R.}~\bibnamefont {Acharya}}, \bibinfo {author} {\bibfnamefont {J.}~\bibnamefont {Iveland}}, \bibinfo {author} {\bibfnamefont {W.}~\bibnamefont {Liu}}, \bibinfo {author} {\bibfnamefont {S.}~\bibnamefont {Kim}}, \bibinfo {author} {\bibfnamefont {B.}~\bibnamefont {Burkett}}, \bibinfo {author} {\bibfnamefont {A.}~\bibnamefont {Megrant}}, \bibinfo {author} {\bibfnamefont {Y.}~\bibnamefont {Chen}}, \bibinfo {author} {\bibfnamefont {C.}~\bibnamefont {Neill}}, \bibinfo {author} {\bibfnamefont {D.}~\bibnamefont {Sank}}, \bibinfo {author} {\bibfnamefont {M.}~\bibnamefont {Devoret}},\ and\ \bibinfo {author} {\bibfnamefont {A.}~\bibnamefont {Opremcak}},\ }\bibfield  {title} {\bibinfo {title} {{Resisting High-Energy Impact Events through Gap Engineering in Superconducting Qubit Arrays}},\ }\href {https://doi.org/10.1103/PhysRevLett.133.240601} {\bibfield  {journal} {\bibinfo  {journal} {Phys. Rev. Lett.}\ }\textbf {\bibinfo {volume} {133}},\ \bibinfo {pages} {240601}
  (\bibinfo {year} {2024})}\BibitemShut {NoStop}%
\bibitem [{\citenamefont {Connolly}\ \emph {et~al.}(2024)\citenamefont {Connolly}, \citenamefont {Kurilovich}, \citenamefont {Diamond}, \citenamefont {Nho}, \citenamefont {B{\o}ttcher}, \citenamefont {Glazman}, \citenamefont {Fatemi},\ and\ \citenamefont {Devoret}}]{Connolly2024May}%
  \BibitemOpen
  \bibfield  {author} {\bibinfo {author} {\bibfnamefont {T.}~\bibnamefont {Connolly}}, \bibinfo {author} {\bibfnamefont {P.~D.}\ \bibnamefont {Kurilovich}}, \bibinfo {author} {\bibfnamefont {S.}~\bibnamefont {Diamond}}, \bibinfo {author} {\bibfnamefont {H.}~\bibnamefont {Nho}}, \bibinfo {author} {\bibfnamefont {C.~G.~L.}\ \bibnamefont {B{\o}ttcher}}, \bibinfo {author} {\bibfnamefont {L.~I.}\ \bibnamefont {Glazman}}, \bibinfo {author} {\bibfnamefont {V.}~\bibnamefont {Fatemi}},\ and\ \bibinfo {author} {\bibfnamefont {M.~H.}\ \bibnamefont {Devoret}},\ }\bibfield  {title} {\bibinfo {title} {{Coexistence of Nonequilibrium Density and Equilibrium Energy Distribution of Quasiparticles in a Superconducting Qubit}},\ }\href {https://doi.org/10.1103/PhysRevLett.132.217001} {\bibfield  {journal} {\bibinfo  {journal} {Phys. Rev. Lett.}\ }\textbf {\bibinfo {volume} {132}},\ \bibinfo {pages} {217001} (\bibinfo {year} {2024})}\BibitemShut {NoStop}%
\bibitem [{\citenamefont {Krause}\ \emph {et~al.}(2024)\citenamefont {Krause}, \citenamefont {Marchegiani}, \citenamefont {Janssen}, \citenamefont {Catelani}, \citenamefont {Ando},\ and\ \citenamefont {Dickel}}]{Krause2024Mar}%
  \BibitemOpen
  \bibfield  {author} {\bibinfo {author} {\bibfnamefont {J.}~\bibnamefont {Krause}}, \bibinfo {author} {\bibfnamefont {G.}~\bibnamefont {Marchegiani}}, \bibinfo {author} {\bibfnamefont {L.~M.}\ \bibnamefont {Janssen}}, \bibinfo {author} {\bibfnamefont {G.}~\bibnamefont {Catelani}}, \bibinfo {author} {\bibfnamefont {Y.}~\bibnamefont {Ando}},\ and\ \bibinfo {author} {\bibfnamefont {C.}~\bibnamefont {Dickel}},\ }\bibfield  {title} {\bibinfo {title} {{Quasiparticle effects in magnetic-field-resilient three-dimensional transmons}},\ }\href {https://doi.org/10.1103/PhysRevApplied.22.044063} {\bibfield  {journal} {\bibinfo  {journal} {Phys. Rev. Appl.}\ }\textbf {\bibinfo {volume} {22}},\ \bibinfo {pages} {044063} (\bibinfo {year} {2024})}\BibitemShut {NoStop}%
\bibitem [{\citenamefont {Christensen}\ \emph {et~al.}(2019)\citenamefont {Christensen}, \citenamefont {Wilen}, \citenamefont {Opremcak}, \citenamefont {Nelson}, \citenamefont {Schlenker}, \citenamefont {Zimonick}, \citenamefont {Faoro}, \citenamefont {Ioffe}, \citenamefont {Rosen}, \citenamefont {DuBois}, \citenamefont {Plourde},\ and\ \citenamefont {McDermott}}]{Christensen2019Oct}%
  \BibitemOpen
  \bibfield  {author} {\bibinfo {author} {\bibfnamefont {B.~G.}\ \bibnamefont {Christensen}}, \bibinfo {author} {\bibfnamefont {C.~D.}\ \bibnamefont {Wilen}}, \bibinfo {author} {\bibfnamefont {A.}~\bibnamefont {Opremcak}}, \bibinfo {author} {\bibfnamefont {J.}~\bibnamefont {Nelson}}, \bibinfo {author} {\bibfnamefont {F.}~\bibnamefont {Schlenker}}, \bibinfo {author} {\bibfnamefont {C.~H.}\ \bibnamefont {Zimonick}}, \bibinfo {author} {\bibfnamefont {L.}~\bibnamefont {Faoro}}, \bibinfo {author} {\bibfnamefont {L.~B.}\ \bibnamefont {Ioffe}}, \bibinfo {author} {\bibfnamefont {Y.~J.}\ \bibnamefont {Rosen}}, \bibinfo {author} {\bibfnamefont {J.~L.}\ \bibnamefont {DuBois}}, \bibinfo {author} {\bibfnamefont {B.~L.~T.}\ \bibnamefont {Plourde}},\ and\ \bibinfo {author} {\bibfnamefont {R.}~\bibnamefont {McDermott}},\ }\bibfield  {title} {\bibinfo {title} {{Anomalous charge noise in superconducting qubits}},\ }\href {https://doi.org/10.1103/PhysRevB.100.140503} {\bibfield  {journal} {\bibinfo  {journal} {Phys. Rev. B}\
  }\textbf {\bibinfo {volume} {100}},\ \bibinfo {pages} {140503} (\bibinfo {year} {2019})}\BibitemShut {NoStop}%
\bibitem [{\citenamefont {Kumar}\ \emph {et~al.}(2016)\citenamefont {Kumar}, \citenamefont {Sendelbach}, \citenamefont {Beck}, \citenamefont {Freeland}, \citenamefont {Wang}, \citenamefont {Wang}, \citenamefont {Yu}, \citenamefont {Wu}, \citenamefont {Pappas},\ and\ \citenamefont {McDermott}}]{Kumar2016Oct}%
  \BibitemOpen
  \bibfield  {author} {\bibinfo {author} {\bibfnamefont {P.}~\bibnamefont {Kumar}}, \bibinfo {author} {\bibfnamefont {S.}~\bibnamefont {Sendelbach}}, \bibinfo {author} {\bibfnamefont {M.~A.}\ \bibnamefont {Beck}}, \bibinfo {author} {\bibfnamefont {J.~W.}\ \bibnamefont {Freeland}}, \bibinfo {author} {\bibfnamefont {Z.}~\bibnamefont {Wang}}, \bibinfo {author} {\bibfnamefont {H.}~\bibnamefont {Wang}}, \bibinfo {author} {\bibfnamefont {C.~C.}\ \bibnamefont {Yu}}, \bibinfo {author} {\bibfnamefont {R.~Q.}\ \bibnamefont {Wu}}, \bibinfo {author} {\bibfnamefont {D.~P.}\ \bibnamefont {Pappas}},\ and\ \bibinfo {author} {\bibfnamefont {R.}~\bibnamefont {McDermott}},\ }\bibfield  {title} {\bibinfo {title} {{Origin and Reduction of $1/f$ Magnetic Flux Noise in Superconducting Devices}},\ }\href {https://doi.org/10.1103/PhysRevApplied.6.041001} {\bibfield  {journal} {\bibinfo  {journal} {Phys. Rev. Appl.}\ }\textbf {\bibinfo {volume} {6}},\ \bibinfo {pages} {041001} (\bibinfo {year} {2016})}\BibitemShut {NoStop}%
\bibitem [{\citenamefont {Anton}\ \emph {et~al.}(2013)\citenamefont {Anton}, \citenamefont {Birenbaum}, \citenamefont {O{'}Kelley}, \citenamefont {Bolkhovsky}, \citenamefont {Braje}, \citenamefont {Fitch}, \citenamefont {Neeley}, \citenamefont {Hilton}, \citenamefont {Cho}, \citenamefont {Irwin}, \citenamefont {Wellstood}, \citenamefont {Oliver}, \citenamefont {Shnirman},\ and\ \citenamefont {Clarke}}]{Anton2013Apr}%
  \BibitemOpen
  \bibfield  {author} {\bibinfo {author} {\bibfnamefont {S.~M.}\ \bibnamefont {Anton}}, \bibinfo {author} {\bibfnamefont {J.~S.}\ \bibnamefont {Birenbaum}}, \bibinfo {author} {\bibfnamefont {S.~R.}\ \bibnamefont {O{'}Kelley}}, \bibinfo {author} {\bibfnamefont {V.}~\bibnamefont {Bolkhovsky}}, \bibinfo {author} {\bibfnamefont {D.~A.}\ \bibnamefont {Braje}}, \bibinfo {author} {\bibfnamefont {G.}~\bibnamefont {Fitch}}, \bibinfo {author} {\bibfnamefont {M.}~\bibnamefont {Neeley}}, \bibinfo {author} {\bibfnamefont {G.~C.}\ \bibnamefont {Hilton}}, \bibinfo {author} {\bibfnamefont {H.-M.}\ \bibnamefont {Cho}}, \bibinfo {author} {\bibfnamefont {K.~D.}\ \bibnamefont {Irwin}}, \bibinfo {author} {\bibfnamefont {F.~C.}\ \bibnamefont {Wellstood}}, \bibinfo {author} {\bibfnamefont {W.~D.}\ \bibnamefont {Oliver}}, \bibinfo {author} {\bibfnamefont {A.}~\bibnamefont {Shnirman}},\ and\ \bibinfo {author} {\bibfnamefont {J.}~\bibnamefont {Clarke}},\ }\bibfield  {title} {\bibinfo {title} {{Magnetic Flux Noise in dc SQUIDs:
  Temperature and Geometry Dependence}},\ }\href {https://doi.org/10.1103/PhysRevLett.110.147002} {\bibfield  {journal} {\bibinfo  {journal} {Phys. Rev. Lett.}\ }\textbf {\bibinfo {volume} {110}},\ \bibinfo {pages} {147002} (\bibinfo {year} {2013})}\BibitemShut {NoStop}%
\bibitem [{\citenamefont {Braum{\ifmmode\ddot{u}\else\"{u}\fi}ller}\ \emph {et~al.}(2020)\citenamefont {Braum{\ifmmode\ddot{u}\else\"{u}\fi}ller}, \citenamefont {Ding}, \citenamefont {Veps{\ifmmode\ddot{a}\else\"{a}\fi}l{\ifmmode\ddot{a}\else\"{a}\fi}inen}, \citenamefont {Sung}, \citenamefont {Kjaergaard}, \citenamefont {Menke}, \citenamefont {Winik}, \citenamefont {Kim}, \citenamefont {Niedzielski}, \citenamefont {Melville}, \citenamefont {Yoder}, \citenamefont {Hirjibehedin}, \citenamefont {Orlando}, \citenamefont {Gustavsson},\ and\ \citenamefont {Oliver}}]{Braumuller2020May}%
  \BibitemOpen
  \bibfield  {author} {\bibinfo {author} {\bibfnamefont {J.}~\bibnamefont {Braum{\ifmmode\ddot{u}\else\"{u}\fi}ller}}, \bibinfo {author} {\bibfnamefont {L.}~\bibnamefont {Ding}}, \bibinfo {author} {\bibfnamefont {A.~P.}\ \bibnamefont {Veps{\ifmmode\ddot{a}\else\"{a}\fi}l{\ifmmode\ddot{a}\else\"{a}\fi}inen}}, \bibinfo {author} {\bibfnamefont {Y.}~\bibnamefont {Sung}}, \bibinfo {author} {\bibfnamefont {M.}~\bibnamefont {Kjaergaard}}, \bibinfo {author} {\bibfnamefont {T.}~\bibnamefont {Menke}}, \bibinfo {author} {\bibfnamefont {R.}~\bibnamefont {Winik}}, \bibinfo {author} {\bibfnamefont {D.}~\bibnamefont {Kim}}, \bibinfo {author} {\bibfnamefont {B.~M.}\ \bibnamefont {Niedzielski}}, \bibinfo {author} {\bibfnamefont {A.}~\bibnamefont {Melville}}, \bibinfo {author} {\bibfnamefont {J.~L.}\ \bibnamefont {Yoder}}, \bibinfo {author} {\bibfnamefont {C.~F.}\ \bibnamefont {Hirjibehedin}}, \bibinfo {author} {\bibfnamefont {T.~P.}\ \bibnamefont {Orlando}}, \bibinfo {author} {\bibfnamefont {S.}~\bibnamefont {Gustavsson}},\
  and\ \bibinfo {author} {\bibfnamefont {W.~D.}\ \bibnamefont {Oliver}},\ }\bibfield  {title} {\bibinfo {title} {{Characterizing and Optimizing Qubit Coherence Based on SQUID Geometry}},\ }\href {https://doi.org/10.1103/PhysRevApplied.13.054079} {\bibfield  {journal} {\bibinfo  {journal} {Phys. Rev. Appl.}\ }\textbf {\bibinfo {volume} {13}},\ \bibinfo {pages} {054079} (\bibinfo {year} {2020})}\BibitemShut {NoStop}%
\bibitem [{\citenamefont {Stern}\ \emph {et~al.}(2014)\citenamefont {Stern}, \citenamefont {Catelani}, \citenamefont {Kubo}, \citenamefont {Grezes}, \citenamefont {Bienfait}, \citenamefont {Vion}, \citenamefont {Esteve},\ and\ \citenamefont {Bertet}}]{Stern2014Sep}%
  \BibitemOpen
  \bibfield  {author} {\bibinfo {author} {\bibfnamefont {M.}~\bibnamefont {Stern}}, \bibinfo {author} {\bibfnamefont {G.}~\bibnamefont {Catelani}}, \bibinfo {author} {\bibfnamefont {Y.}~\bibnamefont {Kubo}}, \bibinfo {author} {\bibfnamefont {C.}~\bibnamefont {Grezes}}, \bibinfo {author} {\bibfnamefont {A.}~\bibnamefont {Bienfait}}, \bibinfo {author} {\bibfnamefont {D.}~\bibnamefont {Vion}}, \bibinfo {author} {\bibfnamefont {D.}~\bibnamefont {Esteve}},\ and\ \bibinfo {author} {\bibfnamefont {P.}~\bibnamefont {Bertet}},\ }\bibfield  {title} {\bibinfo {title} {{Flux Qubits with Long Coherence Times for Hybrid Quantum Circuits}},\ }\href {https://doi.org/10.1103/PhysRevLett.113.123601} {\bibfield  {journal} {\bibinfo  {journal} {Phys. Rev. Lett.}\ }\textbf {\bibinfo {volume} {113}},\ \bibinfo {pages} {123601} (\bibinfo {year} {2014})}\BibitemShut {NoStop}%
\bibitem [{\citenamefont {Yoshihara}\ \emph {et~al.}(2006)\citenamefont {Yoshihara}, \citenamefont {Harrabi}, \citenamefont {Niskanen}, \citenamefont {Nakamura},\ and\ \citenamefont {Tsai}}]{Yoshihara2006Oct}%
  \BibitemOpen
  \bibfield  {author} {\bibinfo {author} {\bibfnamefont {F.}~\bibnamefont {Yoshihara}}, \bibinfo {author} {\bibfnamefont {K.}~\bibnamefont {Harrabi}}, \bibinfo {author} {\bibfnamefont {A.~O.}\ \bibnamefont {Niskanen}}, \bibinfo {author} {\bibfnamefont {Y.}~\bibnamefont {Nakamura}},\ and\ \bibinfo {author} {\bibfnamefont {J.~S.}\ \bibnamefont {Tsai}},\ }\bibfield  {title} {\bibinfo {title} {{Decoherence of Flux Qubits due to $1/f$ Flux Noise}},\ }\href {https://doi.org/10.1103/PhysRevLett.97.167001} {\bibfield  {journal} {\bibinfo  {journal} {Phys. Rev. Lett.}\ }\textbf {\bibinfo {volume} {97}},\ \bibinfo {pages} {167001} (\bibinfo {year} {2006})}\BibitemShut {NoStop}%
\bibitem [{\citenamefont {Nugroho}\ \emph {et~al.}(2013)\citenamefont {Nugroho}, \citenamefont {Orlyanchik},\ and\ \citenamefont {Van~Harlingen}}]{Nugroho2013Apr}%
  \BibitemOpen
  \bibfield  {author} {\bibinfo {author} {\bibfnamefont {C.~D.}\ \bibnamefont {Nugroho}}, \bibinfo {author} {\bibfnamefont {V.}~\bibnamefont {Orlyanchik}},\ and\ \bibinfo {author} {\bibfnamefont {D.~J.}\ \bibnamefont {Van~Harlingen}},\ }\bibfield  {title} {\bibinfo {title} {{Low frequency resistance and critical current fluctuations in Al-based Josephson junctions}},\ }\href {https://doi.org/10.1063/1.4801521} {\bibfield  {journal} {\bibinfo  {journal} {Appl. Phys. Lett.}\ }\textbf {\bibinfo {volume} {102}},\ \bibinfo {pages} {142602} (\bibinfo {year} {2013})}\BibitemShut {NoStop}%
\bibitem [{\citenamefont {Yan}\ \emph {et~al.}(2016)\citenamefont {Yan}, \citenamefont {Gustavsson}, \citenamefont {Kamal}, \citenamefont {Birenbaum}, \citenamefont {Sears}, \citenamefont {Hover}, \citenamefont {Gudmundsen}, \citenamefont {Rosenberg}, \citenamefont {Samach}, \citenamefont {Weber}, \citenamefont {Yoder}, \citenamefont {Orlando}, \citenamefont {Clarke}, \citenamefont {Kerman},\ and\ \citenamefont {Oliver}}]{Yan2016Nov}%
  \BibitemOpen
  \bibfield  {author} {\bibinfo {author} {\bibfnamefont {F.}~\bibnamefont {Yan}}, \bibinfo {author} {\bibfnamefont {S.}~\bibnamefont {Gustavsson}}, \bibinfo {author} {\bibfnamefont {A.}~\bibnamefont {Kamal}}, \bibinfo {author} {\bibfnamefont {J.}~\bibnamefont {Birenbaum}}, \bibinfo {author} {\bibfnamefont {A.~P.}\ \bibnamefont {Sears}}, \bibinfo {author} {\bibfnamefont {D.}~\bibnamefont {Hover}}, \bibinfo {author} {\bibfnamefont {T.~J.}\ \bibnamefont {Gudmundsen}}, \bibinfo {author} {\bibfnamefont {D.}~\bibnamefont {Rosenberg}}, \bibinfo {author} {\bibfnamefont {G.}~\bibnamefont {Samach}}, \bibinfo {author} {\bibfnamefont {S.}~\bibnamefont {Weber}}, \bibinfo {author} {\bibfnamefont {J.~L.}\ \bibnamefont {Yoder}}, \bibinfo {author} {\bibfnamefont {T.~P.}\ \bibnamefont {Orlando}}, \bibinfo {author} {\bibfnamefont {J.}~\bibnamefont {Clarke}}, \bibinfo {author} {\bibfnamefont {A.~J.}\ \bibnamefont {Kerman}},\ and\ \bibinfo {author} {\bibfnamefont {W.~D.}\ \bibnamefont {Oliver}},\ }\bibfield  {title} {\bibinfo
  {title} {{The flux qubit revisited to enhance coherence and reproducibility}},\ }\href {https://doi.org/10.1038/ncomms12964} {\bibfield  {journal} {\bibinfo  {journal} {Nat. Commun.}\ }\textbf {\bibinfo {volume} {7}},\ \bibinfo {pages} {1} (\bibinfo {year} {2016})}\BibitemShut {NoStop}%
\bibitem [{\citenamefont {M{\ifmmode\ddot{u}\else\"{u}\fi}ller}\ \emph {et~al.}(2019)\citenamefont {M{\ifmmode\ddot{u}\else\"{u}\fi}ller}, \citenamefont {Cole},\ and\ \citenamefont {Lisenfeld}}]{Muller2019Oct}%
  \BibitemOpen
  \bibfield  {author} {\bibinfo {author} {\bibfnamefont {C.}~\bibnamefont {M{\ifmmode\ddot{u}\else\"{u}\fi}ller}}, \bibinfo {author} {\bibfnamefont {J.~H.}\ \bibnamefont {Cole}},\ and\ \bibinfo {author} {\bibfnamefont {J.}~\bibnamefont {Lisenfeld}},\ }\bibfield  {title} {\bibinfo {title} {{Towards understanding two-level-systems in amorphous solids: insights from quantum circuits}},\ }\href {https://doi.org/10.1088/1361-6633/ab3a7e} {\bibfield  {journal} {\bibinfo  {journal} {Rep. Prog. Phys.}\ }\textbf {\bibinfo {volume} {82}},\ \bibinfo {pages} {124501} (\bibinfo {year} {2019})}\BibitemShut {NoStop}%
\bibitem [{\citenamefont {Thorbeck}\ \emph {et~al.}(2023)\citenamefont {Thorbeck}, \citenamefont {Eddins}, \citenamefont {Lauer}, \citenamefont {McClure},\ and\ \citenamefont {Carroll}}]{Thorbeck2023Jun}%
  \BibitemOpen
  \bibfield  {author} {\bibinfo {author} {\bibfnamefont {T.}~\bibnamefont {Thorbeck}}, \bibinfo {author} {\bibfnamefont {A.}~\bibnamefont {Eddins}}, \bibinfo {author} {\bibfnamefont {I.}~\bibnamefont {Lauer}}, \bibinfo {author} {\bibfnamefont {D.~T.}\ \bibnamefont {McClure}},\ and\ \bibinfo {author} {\bibfnamefont {M.}~\bibnamefont {Carroll}},\ }\bibfield  {title} {\bibinfo {title} {{Two-Level-System Dynamics in a Superconducting Qubit Due to Background Ionizing Radiation}},\ }\href {https://doi.org/10.1103/PRXQuantum.4.020356} {\bibfield  {journal} {\bibinfo  {journal} {PRX Quantum}\ }\textbf {\bibinfo {volume} {4}},\ \bibinfo {pages} {020356} (\bibinfo {year} {2023})}\BibitemShut {NoStop}%
\bibitem [{\citenamefont {Spiecker}\ \emph {et~al.}(2023)\citenamefont {Spiecker}, \citenamefont {Paluch}, \citenamefont {Gosling}, \citenamefont {Drucker}, \citenamefont {Matityahu}, \citenamefont {Gusenkova}, \citenamefont {G{\ifmmode\ddot{u}\else\"{u}\fi}nzler}, \citenamefont {Rieger}, \citenamefont {Takmakov}, \citenamefont {Valenti}, \citenamefont {Winkel}, \citenamefont {Gebauer}, \citenamefont {Sander}, \citenamefont {Catelani}, \citenamefont {Shnirman}, \citenamefont {Ustinov}, \citenamefont {Wernsdorfer}, \citenamefont {Cohen},\ and\ \citenamefont {Pop}}]{Spiecker2023Sep}%
  \BibitemOpen
  \bibfield  {author} {\bibinfo {author} {\bibfnamefont {M.}~\bibnamefont {Spiecker}}, \bibinfo {author} {\bibfnamefont {P.}~\bibnamefont {Paluch}}, \bibinfo {author} {\bibfnamefont {N.}~\bibnamefont {Gosling}}, \bibinfo {author} {\bibfnamefont {N.}~\bibnamefont {Drucker}}, \bibinfo {author} {\bibfnamefont {S.}~\bibnamefont {Matityahu}}, \bibinfo {author} {\bibfnamefont {D.}~\bibnamefont {Gusenkova}}, \bibinfo {author} {\bibfnamefont {S.}~\bibnamefont {G{\ifmmode\ddot{u}\else\"{u}\fi}nzler}}, \bibinfo {author} {\bibfnamefont {D.}~\bibnamefont {Rieger}}, \bibinfo {author} {\bibfnamefont {I.}~\bibnamefont {Takmakov}}, \bibinfo {author} {\bibfnamefont {F.}~\bibnamefont {Valenti}}, \bibinfo {author} {\bibfnamefont {P.}~\bibnamefont {Winkel}}, \bibinfo {author} {\bibfnamefont {R.}~\bibnamefont {Gebauer}}, \bibinfo {author} {\bibfnamefont {O.}~\bibnamefont {Sander}}, \bibinfo {author} {\bibfnamefont {G.}~\bibnamefont {Catelani}}, \bibinfo {author} {\bibfnamefont {A.}~\bibnamefont {Shnirman}}, \bibinfo {author}
  {\bibfnamefont {A.~V.}\ \bibnamefont {Ustinov}}, \bibinfo {author} {\bibfnamefont {W.}~\bibnamefont {Wernsdorfer}}, \bibinfo {author} {\bibfnamefont {Y.}~\bibnamefont {Cohen}},\ and\ \bibinfo {author} {\bibfnamefont {I.~M.}\ \bibnamefont {Pop}},\ }\bibfield  {title} {\bibinfo {title} {{Two-level system hyperpolarization using a quantum Szilard engine}},\ }\href {https://doi.org/10.1038/s41567-023-02082-8} {\bibfield  {journal} {\bibinfo  {journal} {Nat. Phys.}\ }\textbf {\bibinfo {volume} {19}},\ \bibinfo {pages} {1320} (\bibinfo {year} {2023})}\BibitemShut {NoStop}%
\bibitem [{\citenamefont {Odeh}\ \emph {et~al.}(2025)\citenamefont {Odeh}, \citenamefont {Godeneli}, \citenamefont {Li}, \citenamefont {Tangirala}, \citenamefont {Zhou}, \citenamefont {Zhang}, \citenamefont {Zhang},\ and\ \citenamefont {Sipahigil}}]{Odeh2023Dec}%
  \BibitemOpen
  \bibfield  {author} {\bibinfo {author} {\bibfnamefont {M.}~\bibnamefont {Odeh}}, \bibinfo {author} {\bibfnamefont {K.}~\bibnamefont {Godeneli}}, \bibinfo {author} {\bibfnamefont {E.}~\bibnamefont {Li}}, \bibinfo {author} {\bibfnamefont {R.}~\bibnamefont {Tangirala}}, \bibinfo {author} {\bibfnamefont {H.}~\bibnamefont {Zhou}}, \bibinfo {author} {\bibfnamefont {X.}~\bibnamefont {Zhang}}, \bibinfo {author} {\bibfnamefont {Z.-H.}\ \bibnamefont {Zhang}},\ and\ \bibinfo {author} {\bibfnamefont {A.}~\bibnamefont {Sipahigil}},\ }\bibfield  {title} {\bibinfo {title} {{Non-Markovian dynamics of a superconducting qubit in a phononic bandgap}},\ }\href {https://doi.org/10.1038/s41567-024-02740-5} {\bibfield  {journal} {\bibinfo  {journal} {Nat. Phys.}\ ,\ \bibinfo {pages} {1}} (\bibinfo {year} {2025})}\BibitemShut {NoStop}%
\bibitem [{\citenamefont {Winkel}\ \emph {et~al.}(2020)\citenamefont {Winkel}, \citenamefont {Borisov}, \citenamefont {Gr{\ifmmode\ddot{u}\else\"{u}\fi}nhaupt}, \citenamefont {Rieger}, \citenamefont {Spiecker}, \citenamefont {Valenti}, \citenamefont {Ustinov}, \citenamefont {Wernsdorfer},\ and\ \citenamefont {Pop}}]{Winkel2020Aug}%
  \BibitemOpen
  \bibfield  {author} {\bibinfo {author} {\bibfnamefont {P.}~\bibnamefont {Winkel}}, \bibinfo {author} {\bibfnamefont {K.}~\bibnamefont {Borisov}}, \bibinfo {author} {\bibfnamefont {L.}~\bibnamefont {Gr{\ifmmode\ddot{u}\else\"{u}\fi}nhaupt}}, \bibinfo {author} {\bibfnamefont {D.}~\bibnamefont {Rieger}}, \bibinfo {author} {\bibfnamefont {M.}~\bibnamefont {Spiecker}}, \bibinfo {author} {\bibfnamefont {F.}~\bibnamefont {Valenti}}, \bibinfo {author} {\bibfnamefont {A.~V.}\ \bibnamefont {Ustinov}}, \bibinfo {author} {\bibfnamefont {W.}~\bibnamefont {Wernsdorfer}},\ and\ \bibinfo {author} {\bibfnamefont {I.~M.}\ \bibnamefont {Pop}},\ }\bibfield  {title} {\bibinfo {title} {{Implementation of a Transmon Qubit Using Superconducting Granular Aluminum}},\ }\href {https://doi.org/10.1103/PhysRevX.10.031032} {\bibfield  {journal} {\bibinfo  {journal} {Phys. Rev. X}\ }\textbf {\bibinfo {volume} {10}},\ \bibinfo {pages} {031032} (\bibinfo {year} {2020})}\BibitemShut {NoStop}%
\bibitem [{\citenamefont {Krause}\ \emph {et~al.}(2022)\citenamefont {Krause}, \citenamefont {Dickel}, \citenamefont {Vaal}, \citenamefont {Vielmetter}, \citenamefont {Feng}, \citenamefont {Bounds}, \citenamefont {Catelani}, \citenamefont {Fink},\ and\ \citenamefont {Ando}}]{Krause2022Mar}%
  \BibitemOpen
  \bibfield  {author} {\bibinfo {author} {\bibfnamefont {J.}~\bibnamefont {Krause}}, \bibinfo {author} {\bibfnamefont {C.}~\bibnamefont {Dickel}}, \bibinfo {author} {\bibfnamefont {E.}~\bibnamefont {Vaal}}, \bibinfo {author} {\bibfnamefont {M.}~\bibnamefont {Vielmetter}}, \bibinfo {author} {\bibfnamefont {J.}~\bibnamefont {Feng}}, \bibinfo {author} {\bibfnamefont {R.}~\bibnamefont {Bounds}}, \bibinfo {author} {\bibfnamefont {G.}~\bibnamefont {Catelani}}, \bibinfo {author} {\bibfnamefont {J.~M.}\ \bibnamefont {Fink}},\ and\ \bibinfo {author} {\bibfnamefont {Y.}~\bibnamefont {Ando}},\ }\bibfield  {title} {\bibinfo {title} {{Magnetic Field Resilience of Three-Dimensional Transmons with Thin-Film ${\text{Al/AlO}}_{x}/\text{Al}$ Josephson Junctions Approaching 1 T}},\ }\href {https://doi.org/10.1103/PhysRevApplied.17.034032} {\bibfield  {journal} {\bibinfo  {journal} {Phys. Rev. Appl.}\ }\textbf {\bibinfo {volume} {17}},\ \bibinfo {pages} {034032} (\bibinfo {year} {2022})}\BibitemShut {NoStop}%
\bibitem [{\citenamefont {Rower}\ \emph {et~al.}(2023)\citenamefont {Rower}, \citenamefont {Ateshian}, \citenamefont {Li}, \citenamefont {Hays}, \citenamefont {Bluvstein}, \citenamefont {Ding}, \citenamefont {Kannan}, \citenamefont {Almanakly}, \citenamefont {Braum{\ifmmode\ddot{u}\else\"{u}\fi}ller}, \citenamefont {Kim}, \citenamefont {Melville}, \citenamefont {Niedzielski}, \citenamefont {Schwartz}, \citenamefont {Yoder}, \citenamefont {Orlando}, \citenamefont {Wang}, \citenamefont {Gustavsson}, \citenamefont {Grover}, \citenamefont {Serniak}, \citenamefont {Comin},\ and\ \citenamefont {Oliver}}]{Rower2023May}%
  \BibitemOpen
  \bibfield  {author} {\bibinfo {author} {\bibfnamefont {D.~A.}\ \bibnamefont {Rower}}, \bibinfo {author} {\bibfnamefont {L.}~\bibnamefont {Ateshian}}, \bibinfo {author} {\bibfnamefont {L.~H.}\ \bibnamefont {Li}}, \bibinfo {author} {\bibfnamefont {M.}~\bibnamefont {Hays}}, \bibinfo {author} {\bibfnamefont {D.}~\bibnamefont {Bluvstein}}, \bibinfo {author} {\bibfnamefont {L.}~\bibnamefont {Ding}}, \bibinfo {author} {\bibfnamefont {B.}~\bibnamefont {Kannan}}, \bibinfo {author} {\bibfnamefont {A.}~\bibnamefont {Almanakly}}, \bibinfo {author} {\bibfnamefont {J.}~\bibnamefont {Braum{\ifmmode\ddot{u}\else\"{u}\fi}ller}}, \bibinfo {author} {\bibfnamefont {D.~K.}\ \bibnamefont {Kim}}, \bibinfo {author} {\bibfnamefont {A.}~\bibnamefont {Melville}}, \bibinfo {author} {\bibfnamefont {B.~M.}\ \bibnamefont {Niedzielski}}, \bibinfo {author} {\bibfnamefont {M.~E.}\ \bibnamefont {Schwartz}}, \bibinfo {author} {\bibfnamefont {J.~L.}\ \bibnamefont {Yoder}}, \bibinfo {author} {\bibfnamefont {T.~P.}\ \bibnamefont {Orlando}},
  \bibinfo {author} {\bibfnamefont {J.~I.-J.}\ \bibnamefont {Wang}}, \bibinfo {author} {\bibfnamefont {S.}~\bibnamefont {Gustavsson}}, \bibinfo {author} {\bibfnamefont {J.~A.}\ \bibnamefont {Grover}}, \bibinfo {author} {\bibfnamefont {K.}~\bibnamefont {Serniak}}, \bibinfo {author} {\bibfnamefont {R.}~\bibnamefont {Comin}},\ and\ \bibinfo {author} {\bibfnamefont {W.~D.}\ \bibnamefont {Oliver}},\ }\bibfield  {title} {\bibinfo {title} {{Evolution of $1/f$ Flux Noise in Superconducting Qubits with Weak Magnetic Fields}},\ }\href {https://doi.org/10.1103/PhysRevLett.130.220602} {\bibfield  {journal} {\bibinfo  {journal} {Phys. Rev. Lett.}\ }\textbf {\bibinfo {volume} {130}},\ \bibinfo {pages} {220602} (\bibinfo {year} {2023})}\BibitemShut {NoStop}%
\bibitem [{\citenamefont {Borisov}\ \emph {et~al.}(2020)\citenamefont {Borisov}, \citenamefont {Rieger}, \citenamefont {Winkel}, \citenamefont {Henriques}, \citenamefont {Valenti}, \citenamefont {Ionita}, \citenamefont {Wessbecher}, \citenamefont {Spiecker}, \citenamefont {Gusenkova}, \citenamefont {Pop},\ and\ \citenamefont {Wernsdorfer}}]{Borisov2020Sep}%
  \BibitemOpen
  \bibfield  {author} {\bibinfo {author} {\bibfnamefont {K.}~\bibnamefont {Borisov}}, \bibinfo {author} {\bibfnamefont {D.}~\bibnamefont {Rieger}}, \bibinfo {author} {\bibfnamefont {P.}~\bibnamefont {Winkel}}, \bibinfo {author} {\bibfnamefont {F.}~\bibnamefont {Henriques}}, \bibinfo {author} {\bibfnamefont {F.}~\bibnamefont {Valenti}}, \bibinfo {author} {\bibfnamefont {A.}~\bibnamefont {Ionita}}, \bibinfo {author} {\bibfnamefont {M.}~\bibnamefont {Wessbecher}}, \bibinfo {author} {\bibfnamefont {M.}~\bibnamefont {Spiecker}}, \bibinfo {author} {\bibfnamefont {D.}~\bibnamefont {Gusenkova}}, \bibinfo {author} {\bibfnamefont {I.~M.}\ \bibnamefont {Pop}},\ and\ \bibinfo {author} {\bibfnamefont {W.}~\bibnamefont {Wernsdorfer}},\ }\bibfield  {title} {\bibinfo {title} {{Superconducting granular aluminum resonators resilient to magnetic fields up to 1 Tesla}},\ }\bibfield  {journal} {\bibinfo  {journal} {Appl. Phys. Lett.}\ }\textbf {\bibinfo {volume} {117}},\ \href {https://doi.org/10.1063/5.0018012}
  {10.1063/5.0018012} (\bibinfo {year} {2020})\BibitemShut {NoStop}%
\bibitem [{\citenamefont {Samkharadze}\ \emph {et~al.}(2016)\citenamefont {Samkharadze}, \citenamefont {Bruno}, \citenamefont {Scarlino}, \citenamefont {Zheng}, \citenamefont {DiVincenzo}, \citenamefont {DiCarlo},\ and\ \citenamefont {Vandersypen}}]{Samkharadze2016Apr}%
  \BibitemOpen
  \bibfield  {author} {\bibinfo {author} {\bibfnamefont {N.}~\bibnamefont {Samkharadze}}, \bibinfo {author} {\bibfnamefont {A.}~\bibnamefont {Bruno}}, \bibinfo {author} {\bibfnamefont {P.}~\bibnamefont {Scarlino}}, \bibinfo {author} {\bibfnamefont {G.}~\bibnamefont {Zheng}}, \bibinfo {author} {\bibfnamefont {D.~P.}\ \bibnamefont {DiVincenzo}}, \bibinfo {author} {\bibfnamefont {L.}~\bibnamefont {DiCarlo}},\ and\ \bibinfo {author} {\bibfnamefont {L.~M.~K.}\ \bibnamefont {Vandersypen}},\ }\bibfield  {title} {\bibinfo {title} {{High-Kinetic-Inductance Superconducting Nanowire Resonators for Circuit QED in a Magnetic Field}},\ }\href {https://doi.org/10.1103/PhysRevApplied.5.044004} {\bibfield  {journal} {\bibinfo  {journal} {Phys. Rev. Appl.}\ }\textbf {\bibinfo {volume} {5}},\ \bibinfo {pages} {044004} (\bibinfo {year} {2016})}\BibitemShut {NoStop}%
\bibitem [{\citenamefont {Kroll}\ \emph {et~al.}(2019)\citenamefont {Kroll}, \citenamefont {Borsoi}, \citenamefont {van~der Enden}, \citenamefont {Uilhoorn}, \citenamefont {de~Jong}, \citenamefont {Quintero-P{\ifmmode\acute{e}\else\'{e}\fi}rez}, \citenamefont {van Woerkom}, \citenamefont {Bruno}, \citenamefont {Plissard}, \citenamefont {Car}, \citenamefont {Bakkers}, \citenamefont {Cassidy},\ and\ \citenamefont {Kouwenhoven}}]{Kroll2019Jun}%
  \BibitemOpen
  \bibfield  {author} {\bibinfo {author} {\bibfnamefont {J.~G.}\ \bibnamefont {Kroll}}, \bibinfo {author} {\bibfnamefont {F.}~\bibnamefont {Borsoi}}, \bibinfo {author} {\bibfnamefont {K.~L.}\ \bibnamefont {van~der Enden}}, \bibinfo {author} {\bibfnamefont {W.}~\bibnamefont {Uilhoorn}}, \bibinfo {author} {\bibfnamefont {D.}~\bibnamefont {de~Jong}}, \bibinfo {author} {\bibfnamefont {M.}~\bibnamefont {Quintero-P{\ifmmode\acute{e}\else\'{e}\fi}rez}}, \bibinfo {author} {\bibfnamefont {D.~J.}\ \bibnamefont {van Woerkom}}, \bibinfo {author} {\bibfnamefont {A.}~\bibnamefont {Bruno}}, \bibinfo {author} {\bibfnamefont {S.~R.}\ \bibnamefont {Plissard}}, \bibinfo {author} {\bibfnamefont {D.}~\bibnamefont {Car}}, \bibinfo {author} {\bibfnamefont {E.~P. A.~M.}\ \bibnamefont {Bakkers}}, \bibinfo {author} {\bibfnamefont {M.~C.}\ \bibnamefont {Cassidy}},\ and\ \bibinfo {author} {\bibfnamefont {L.~P.}\ \bibnamefont {Kouwenhoven}},\ }\bibfield  {title} {\bibinfo {title} {{Magnetic-Field-Resilient Superconducting
  Coplanar-Waveguide Resonators for Hybrid Circuit Quantum Electrodynamics Experiments}},\ }\href {https://doi.org/10.1103/PhysRevApplied.11.064053} {\bibfield  {journal} {\bibinfo  {journal} {Phys. Rev. Appl.}\ }\textbf {\bibinfo {volume} {11}},\ \bibinfo {pages} {064053} (\bibinfo {year} {2019})}\BibitemShut {NoStop}%
\bibitem [{\citenamefont {Kringh{\o}j}\ \emph {et~al.}(2021)\citenamefont {Kringh{\o}j}, \citenamefont {Larsen}, \citenamefont {Erlandsson}, \citenamefont {Uilhoorn}, \citenamefont {Kroll}, \citenamefont {Hesselberg}, \citenamefont {McNeil}, \citenamefont {Krogstrup}, \citenamefont {Casparis}, \citenamefont {Marcus},\ and\ \citenamefont {Petersson}}]{Kringhoj2021May}%
  \BibitemOpen
  \bibfield  {author} {\bibinfo {author} {\bibfnamefont {A.}~\bibnamefont {Kringh{\o}j}}, \bibinfo {author} {\bibfnamefont {T.~W.}\ \bibnamefont {Larsen}}, \bibinfo {author} {\bibfnamefont {O.}~\bibnamefont {Erlandsson}}, \bibinfo {author} {\bibfnamefont {W.}~\bibnamefont {Uilhoorn}}, \bibinfo {author} {\bibfnamefont {J.~G.}\ \bibnamefont {Kroll}}, \bibinfo {author} {\bibfnamefont {M.}~\bibnamefont {Hesselberg}}, \bibinfo {author} {\bibfnamefont {R.~P.~G.}\ \bibnamefont {McNeil}}, \bibinfo {author} {\bibfnamefont {P.}~\bibnamefont {Krogstrup}}, \bibinfo {author} {\bibfnamefont {L.}~\bibnamefont {Casparis}}, \bibinfo {author} {\bibfnamefont {C.~M.}\ \bibnamefont {Marcus}},\ and\ \bibinfo {author} {\bibfnamefont {K.~D.}\ \bibnamefont {Petersson}},\ }\bibfield  {title} {\bibinfo {title} {{Magnetic-Field-Compatible Superconducting Transmon Qubit}},\ }\href {https://doi.org/10.1103/PhysRevApplied.15.054001} {\bibfield  {journal} {\bibinfo  {journal} {Phys. Rev. Appl.}\ }\textbf {\bibinfo {volume} {15}},\ \bibinfo
  {pages} {054001} (\bibinfo {year} {2021})}\BibitemShut {NoStop}%
\bibitem [{\citenamefont {Pita-Vidal}\ \emph {et~al.}(2020)\citenamefont {Pita-Vidal}, \citenamefont {Bargerbos}, \citenamefont {Yang}, \citenamefont {van Woerkom}, \citenamefont {Pfaff}, \citenamefont {Haider}, \citenamefont {Krogstrup}, \citenamefont {Kouwenhoven}, \citenamefont {de~Lange},\ and\ \citenamefont {Kou}}]{Pita-Vidal2020Dec}%
  \BibitemOpen
  \bibfield  {author} {\bibinfo {author} {\bibfnamefont {M.}~\bibnamefont {Pita-Vidal}}, \bibinfo {author} {\bibfnamefont {A.}~\bibnamefont {Bargerbos}}, \bibinfo {author} {\bibfnamefont {C.-K.}\ \bibnamefont {Yang}}, \bibinfo {author} {\bibfnamefont {D.~J.}\ \bibnamefont {van Woerkom}}, \bibinfo {author} {\bibfnamefont {W.}~\bibnamefont {Pfaff}}, \bibinfo {author} {\bibfnamefont {N.}~\bibnamefont {Haider}}, \bibinfo {author} {\bibfnamefont {P.}~\bibnamefont {Krogstrup}}, \bibinfo {author} {\bibfnamefont {L.~P.}\ \bibnamefont {Kouwenhoven}}, \bibinfo {author} {\bibfnamefont {G.}~\bibnamefont {de~Lange}},\ and\ \bibinfo {author} {\bibfnamefont {A.}~\bibnamefont {Kou}},\ }\bibfield  {title} {\bibinfo {title} {{Gate-Tunable Field-Compatible Fluxonium}},\ }\href {https://doi.org/10.1103/PhysRevApplied.14.064038} {\bibfield  {journal} {\bibinfo  {journal} {Phys. Rev. Appl.}\ }\textbf {\bibinfo {volume} {14}},\ \bibinfo {pages} {064038} (\bibinfo {year} {2020})}\BibitemShut {NoStop}%
\bibitem [{\citenamefont {Kroll}\ \emph {et~al.}(2018)\citenamefont {Kroll}, \citenamefont {Uilhoorn}, \citenamefont {van~der Enden}, \citenamefont {de~Jong}, \citenamefont {Watanabe}, \citenamefont {Taniguchi}, \citenamefont {Goswami}, \citenamefont {Cassidy},\ and\ \citenamefont {Kouwenhoven}}]{Kroll2018Nov}%
  \BibitemOpen
  \bibfield  {author} {\bibinfo {author} {\bibfnamefont {J.~G.}\ \bibnamefont {Kroll}}, \bibinfo {author} {\bibfnamefont {W.}~\bibnamefont {Uilhoorn}}, \bibinfo {author} {\bibfnamefont {K.~L.}\ \bibnamefont {van~der Enden}}, \bibinfo {author} {\bibfnamefont {D.}~\bibnamefont {de~Jong}}, \bibinfo {author} {\bibfnamefont {K.}~\bibnamefont {Watanabe}}, \bibinfo {author} {\bibfnamefont {T.}~\bibnamefont {Taniguchi}}, \bibinfo {author} {\bibfnamefont {S.}~\bibnamefont {Goswami}}, \bibinfo {author} {\bibfnamefont {M.~C.}\ \bibnamefont {Cassidy}},\ and\ \bibinfo {author} {\bibfnamefont {L.~P.}\ \bibnamefont {Kouwenhoven}},\ }\bibfield  {title} {\bibinfo {title} {{Magnetic field compatible circuit quantum electrodynamics with graphene Josephson junctions}},\ }\href {https://doi.org/10.1038/s41467-018-07124-x} {\bibfield  {journal} {\bibinfo  {journal} {Nat. Commun.}\ }\textbf {\bibinfo {volume} {9}},\ \bibinfo {pages} {1} (\bibinfo {year} {2018})}\BibitemShut {NoStop}%
\bibitem [{\citenamefont {Deutscher}\ and\ \citenamefont {Dodds}(1977)}]{Deutscher1977Nov}%
  \BibitemOpen
  \bibfield  {author} {\bibinfo {author} {\bibfnamefont {G.}~\bibnamefont {Deutscher}}\ and\ \bibinfo {author} {\bibfnamefont {S.~A.}\ \bibnamefont {Dodds}},\ }\bibfield  {title} {\bibinfo {title} {{Critical-field anisotropy and fluctuation conductivity in granular aluminum films}},\ }\href {https://doi.org/10.1103/PhysRevB.16.3936} {\bibfield  {journal} {\bibinfo  {journal} {Phys. Rev. B}\ }\textbf {\bibinfo {volume} {16}},\ \bibinfo {pages} {3936} (\bibinfo {year} {1977})}\BibitemShut {NoStop}%
\bibitem [{\citenamefont {Rieger}\ \emph {et~al.}(2023{\natexlab{a}})\citenamefont {Rieger}, \citenamefont {G{\ifmmode\ddot{u}\else\"{u}\fi}nzler}, \citenamefont {Spiecker}, \citenamefont {Paluch}, \citenamefont {Winkel}, \citenamefont {Hahn}, \citenamefont {Hohmann}, \citenamefont {Bacher}, \citenamefont {Wernsdorfer},\ and\ \citenamefont {Pop}}]{gralmonium}%
  \BibitemOpen
  \bibfield  {author} {\bibinfo {author} {\bibfnamefont {D.}~\bibnamefont {Rieger}}, \bibinfo {author} {\bibfnamefont {S.}~\bibnamefont {G{\ifmmode\ddot{u}\else\"{u}\fi}nzler}}, \bibinfo {author} {\bibfnamefont {M.}~\bibnamefont {Spiecker}}, \bibinfo {author} {\bibfnamefont {P.}~\bibnamefont {Paluch}}, \bibinfo {author} {\bibfnamefont {P.}~\bibnamefont {Winkel}}, \bibinfo {author} {\bibfnamefont {L.}~\bibnamefont {Hahn}}, \bibinfo {author} {\bibfnamefont {J.~K.}\ \bibnamefont {Hohmann}}, \bibinfo {author} {\bibfnamefont {A.}~\bibnamefont {Bacher}}, \bibinfo {author} {\bibfnamefont {W.}~\bibnamefont {Wernsdorfer}},\ and\ \bibinfo {author} {\bibfnamefont {I.~M.}\ \bibnamefont {Pop}},\ }\bibfield  {title} {\bibinfo {title} {{Granular aluminium nanojunction fluxonium qubit}},\ }\href {https://doi.org/10.1038/s41563-022-01417-9} {\bibfield  {journal} {\bibinfo  {journal} {Nat. Mater.}\ }\textbf {\bibinfo {volume} {22}},\ \bibinfo {pages} {194} (\bibinfo {year} {2023}{\natexlab{a}})}\BibitemShut {NoStop}%
\bibitem [{\citenamefont {Gusenkova}\ \emph {et~al.}(2022)\citenamefont {Gusenkova}, \citenamefont {Valenti}, \citenamefont {Spiecker}, \citenamefont {G{\ifmmode\ddot{u}\else\"{u}\fi}nzler}, \citenamefont {Paluch}, \citenamefont {Rieger}, \citenamefont {Piora{\ifmmode\mbox{\c{s}}\else\c{s}\fi}-{\ifmmode\mbox{\c{T}}\else\c{T}\fi}imbolma{\ifmmode\mbox{\c{s}}\else\c{s}\fi}}, \citenamefont {Z{\ifmmode\hat{a}\else\^{a}\fi}rbo}, \citenamefont {Casali}, \citenamefont {Colantoni}, \citenamefont {Cruciani}, \citenamefont {Pirro}, \citenamefont {Cardani}, \citenamefont {Petrescu}, \citenamefont {Wernsdorfer}, \citenamefont {Winkel},\ and\ \citenamefont {Pop}}]{Gusenkova2022Jan}%
  \BibitemOpen
  \bibfield  {author} {\bibinfo {author} {\bibfnamefont {D.}~\bibnamefont {Gusenkova}}, \bibinfo {author} {\bibfnamefont {F.}~\bibnamefont {Valenti}}, \bibinfo {author} {\bibfnamefont {M.}~\bibnamefont {Spiecker}}, \bibinfo {author} {\bibfnamefont {S.}~\bibnamefont {G{\ifmmode\ddot{u}\else\"{u}\fi}nzler}}, \bibinfo {author} {\bibfnamefont {P.}~\bibnamefont {Paluch}}, \bibinfo {author} {\bibfnamefont {D.}~\bibnamefont {Rieger}}, \bibinfo {author} {\bibfnamefont {L.-M.}\ \bibnamefont {Piora{\ifmmode\mbox{\c{s}}\else\c{s}\fi}-{\ifmmode\mbox{\c{T}}\else\c{T}\fi}imbolma{\ifmmode\mbox{\c{s}}\else\c{s}\fi}}}, \bibinfo {author} {\bibfnamefont {L.~P.}\ \bibnamefont {Z{\ifmmode\hat{a}\else\^{a}\fi}rbo}}, \bibinfo {author} {\bibfnamefont {N.}~\bibnamefont {Casali}}, \bibinfo {author} {\bibfnamefont {I.}~\bibnamefont {Colantoni}}, \bibinfo {author} {\bibfnamefont {A.}~\bibnamefont {Cruciani}}, \bibinfo {author} {\bibfnamefont {S.}~\bibnamefont {Pirro}}, \bibinfo {author} {\bibfnamefont {L.}~\bibnamefont {Cardani}},
  \bibinfo {author} {\bibfnamefont {A.}~\bibnamefont {Petrescu}}, \bibinfo {author} {\bibfnamefont {W.}~\bibnamefont {Wernsdorfer}}, \bibinfo {author} {\bibfnamefont {P.}~\bibnamefont {Winkel}},\ and\ \bibinfo {author} {\bibfnamefont {I.~M.}\ \bibnamefont {Pop}},\ }\bibfield  {title} {\bibinfo {title} {{Operating in a deep underground facility improves the locking of gradiometric fluxonium qubits at the sweet spots}},\ }\bibfield  {journal} {\bibinfo  {journal} {Appl. Phys. Lett.}\ }\textbf {\bibinfo {volume} {120}},\ \href {https://doi.org/10.1063/5.0075909} {10.1063/5.0075909} (\bibinfo {year} {2022})\BibitemShut {NoStop}%
\bibitem [{\citenamefont {Zhuang}\ \emph {et~al.}(2025)\citenamefont {Zhuang}, \citenamefont {Rosenstock}, \citenamefont {Liu}, \citenamefont {Somoroff}, \citenamefont {Manucharyan},\ and\ \citenamefont {Wang}}]{Zhuang2025Mar}%
  \BibitemOpen
  \bibfield  {author} {\bibinfo {author} {\bibfnamefont {Z.-T.}\ \bibnamefont {Zhuang}}, \bibinfo {author} {\bibfnamefont {D.}~\bibnamefont {Rosenstock}}, \bibinfo {author} {\bibfnamefont {B.-J.}\ \bibnamefont {Liu}}, \bibinfo {author} {\bibfnamefont {A.}~\bibnamefont {Somoroff}}, \bibinfo {author} {\bibfnamefont {V.~E.}\ \bibnamefont {Manucharyan}},\ and\ \bibinfo {author} {\bibfnamefont {C.}~\bibnamefont {Wang}},\ }\bibfield  {title} {\bibinfo {title} {{Non-Markovian Relaxation Spectroscopy of Fluxonium Qubits}},\ }\bibfield  {journal} {\bibinfo  {journal} {arXiv}\ }\href {https://doi.org/10.48550/arXiv.2503.16381} {10.48550/arXiv.2503.16381} (\bibinfo {year} {2025}),\ \Eprint {https://arxiv.org/abs/2503.16381} {2503.16381} \BibitemShut {NoStop}%
\bibitem [{\citenamefont {Douglass}(1961)}]{Douglass1961Apr}%
  \BibitemOpen
  \bibfield  {author} {\bibinfo {author} {\bibfnamefont {D.~H.}\ \bibnamefont {Douglass}},\ }\bibfield  {title} {\bibinfo {title} {{Magnetic Field Dependence of the Superconducting Energy Gap}},\ }\href {https://doi.org/10.1103/PhysRevLett.6.346} {\bibfield  {journal} {\bibinfo  {journal} {Phys. Rev. Lett.}\ }\textbf {\bibinfo {volume} {6}},\ \bibinfo {pages} {346} (\bibinfo {year} {1961})}\BibitemShut {NoStop}%
\bibitem [{\citenamefont {Deshpande}\ \emph {et~al.}(2025)\citenamefont {Deshpande}, \citenamefont {Pusskeiler}, \citenamefont {Prange}, \citenamefont {Rogge}, \citenamefont {Dressel},\ and\ \citenamefont {Scheffler}}]{Deshpande2024Aug}%
  \BibitemOpen
  \bibfield  {author} {\bibinfo {author} {\bibfnamefont {A.}~\bibnamefont {Deshpande}}, \bibinfo {author} {\bibfnamefont {J.}~\bibnamefont {Pusskeiler}}, \bibinfo {author} {\bibfnamefont {C.}~\bibnamefont {Prange}}, \bibinfo {author} {\bibfnamefont {U.}~\bibnamefont {Rogge}}, \bibinfo {author} {\bibfnamefont {M.}~\bibnamefont {Dressel}},\ and\ \bibinfo {author} {\bibfnamefont {M.}~\bibnamefont {Scheffler}},\ }\bibfield  {title} {\bibinfo {title} {{Tuning the superconducting dome in granular aluminum thin films}},\ }\bibfield  {journal} {\bibinfo  {journal} {J. Appl. Phys.}\ }\textbf {\bibinfo {volume} {137}},\ \href {https://doi.org/10.1063/5.0250146} {10.1063/5.0250146} (\bibinfo {year} {2025})\BibitemShut {NoStop}%
\bibitem [{\citenamefont {Spiecker}\ \emph {et~al.}(2024)\citenamefont {Spiecker}, \citenamefont {Pavlov}, \citenamefont {Shnirman},\ and\ \citenamefont {Pop}}]{Spiecker2024May}%
  \BibitemOpen
  \bibfield  {author} {\bibinfo {author} {\bibfnamefont {M.}~\bibnamefont {Spiecker}}, \bibinfo {author} {\bibfnamefont {A.~I.}\ \bibnamefont {Pavlov}}, \bibinfo {author} {\bibfnamefont {A.}~\bibnamefont {Shnirman}},\ and\ \bibinfo {author} {\bibfnamefont {I.~M.}\ \bibnamefont {Pop}},\ }\bibfield  {title} {\bibinfo {title} {{Solomon equations for qubit and two-level systems: Insights into non-Poissonian quantum jumps}},\ }\href {https://doi.org/10.1103/PhysRevA.109.052218} {\bibfield  {journal} {\bibinfo  {journal} {Phys. Rev. A}\ }\textbf {\bibinfo {volume} {109}},\ \bibinfo {pages} {052218} (\bibinfo {year} {2024})}\BibitemShut {NoStop}%
\bibitem [{\citenamefont {Schwarz}\ \emph {et~al.}(2013)\citenamefont {Schwarz}, \citenamefont {Nagel}, \citenamefont {W{\ifmmode\ddot{o}\else\"{o}\fi}lbing}, \citenamefont {Kemmler}, \citenamefont {Kleiner},\ and\ \citenamefont {Koelle}}]{Schwarz2013Jan}%
  \BibitemOpen
  \bibfield  {author} {\bibinfo {author} {\bibfnamefont {T.}~\bibnamefont {Schwarz}}, \bibinfo {author} {\bibfnamefont {J.}~\bibnamefont {Nagel}}, \bibinfo {author} {\bibfnamefont {R.}~\bibnamefont {W{\ifmmode\ddot{o}\else\"{o}\fi}lbing}}, \bibinfo {author} {\bibfnamefont {M.}~\bibnamefont {Kemmler}}, \bibinfo {author} {\bibfnamefont {R.}~\bibnamefont {Kleiner}},\ and\ \bibinfo {author} {\bibfnamefont {D.}~\bibnamefont {Koelle}},\ }\bibfield  {title} {\bibinfo {title} {{Low-Noise Nano Superconducting Quantum Interference Device Operating in Tesla Magnetic Fields}},\ }\href {https://doi.org/10.1021/nn305431c} {\bibfield  {journal} {\bibinfo  {journal} {ACS Nano}\ }\textbf {\bibinfo {volume} {7}},\ \bibinfo {pages} {844} (\bibinfo {year} {2013})}\BibitemShut {NoStop}%
\bibitem [{\citenamefont {Van~Harlingen}\ \emph {et~al.}(2004)\citenamefont {Van~Harlingen}, \citenamefont {Robertson}, \citenamefont {Plourde}, \citenamefont {Reichardt}, \citenamefont {Crane},\ and\ \citenamefont {Clarke}}]{VanHarlingen2004Aug}%
  \BibitemOpen
  \bibfield  {author} {\bibinfo {author} {\bibfnamefont {D.~J.}\ \bibnamefont {Van~Harlingen}}, \bibinfo {author} {\bibfnamefont {T.~L.}\ \bibnamefont {Robertson}}, \bibinfo {author} {\bibfnamefont {B.~L.~T.}\ \bibnamefont {Plourde}}, \bibinfo {author} {\bibfnamefont {P.~A.}\ \bibnamefont {Reichardt}}, \bibinfo {author} {\bibfnamefont {T.~A.}\ \bibnamefont {Crane}},\ and\ \bibinfo {author} {\bibfnamefont {J.}~\bibnamefont {Clarke}},\ }\bibfield  {title} {\bibinfo {title} {{Decoherence in Josephson-junction qubits due to critical-current fluctuations}},\ }\href {https://doi.org/10.1103/PhysRevB.70.064517} {\bibfield  {journal} {\bibinfo  {journal} {Phys. Rev. B}\ }\textbf {\bibinfo {volume} {70}},\ \bibinfo {pages} {064517} (\bibinfo {year} {2004})}\BibitemShut {NoStop}%
\bibitem [{\citenamefont {Kristen}\ \emph {et~al.}(2024)\citenamefont {Kristen}, \citenamefont {Voss}, \citenamefont {Wildermuth}, \citenamefont {Bilmes}, \citenamefont {Lisenfeld}, \citenamefont {Rotzinger},\ and\ \citenamefont {Ustinov}}]{Kristen2024May}%
  \BibitemOpen
  \bibfield  {author} {\bibinfo {author} {\bibfnamefont {M.}~\bibnamefont {Kristen}}, \bibinfo {author} {\bibfnamefont {J.~N.}\ \bibnamefont {Voss}}, \bibinfo {author} {\bibfnamefont {M.}~\bibnamefont {Wildermuth}}, \bibinfo {author} {\bibfnamefont {A.}~\bibnamefont {Bilmes}}, \bibinfo {author} {\bibfnamefont {J.}~\bibnamefont {Lisenfeld}}, \bibinfo {author} {\bibfnamefont {H.}~\bibnamefont {Rotzinger}},\ and\ \bibinfo {author} {\bibfnamefont {A.~V.}\ \bibnamefont {Ustinov}},\ }\bibfield  {title} {\bibinfo {title} {{Giant Two-Level Systems in a Granular Superconductor}},\ }\href {https://doi.org/10.1103/PhysRevLett.132.217002} {\bibfield  {journal} {\bibinfo  {journal} {Phys. Rev. Lett.}\ }\textbf {\bibinfo {volume} {132}},\ \bibinfo {pages} {217002} (\bibinfo {year} {2024})}\BibitemShut {NoStop}%
\bibitem [{\citenamefont {Abdurakhimov}\ \emph {et~al.}(2022)\citenamefont {Abdurakhimov}, \citenamefont {Mahboob}, \citenamefont {Toida}, \citenamefont {Kakuyanagi}, \citenamefont {Matsuzaki},\ and\ \citenamefont {Saito}}]{Abdurakhimov2022Dec}%
  \BibitemOpen
  \bibfield  {author} {\bibinfo {author} {\bibfnamefont {L.~V.}\ \bibnamefont {Abdurakhimov}}, \bibinfo {author} {\bibfnamefont {I.}~\bibnamefont {Mahboob}}, \bibinfo {author} {\bibfnamefont {H.}~\bibnamefont {Toida}}, \bibinfo {author} {\bibfnamefont {K.}~\bibnamefont {Kakuyanagi}}, \bibinfo {author} {\bibfnamefont {Y.}~\bibnamefont {Matsuzaki}},\ and\ \bibinfo {author} {\bibfnamefont {S.}~\bibnamefont {Saito}},\ }\bibfield  {title} {\bibinfo {title} {{Identification of Different Types of High-Frequency Defects in Superconducting Qubits}},\ }\href {https://doi.org/10.1103/PRXQuantum.3.040332} {\bibfield  {journal} {\bibinfo  {journal} {PRX Quantum}\ }\textbf {\bibinfo {volume} {3}},\ \bibinfo {pages} {040332} (\bibinfo {year} {2022})}\BibitemShut {NoStop}%
\bibitem [{\citenamefont {Schriefl}\ \emph {et~al.}(2006)\citenamefont {Schriefl}, \citenamefont {Makhlin}, \citenamefont {Shnirman},\ and\ \citenamefont {Sch{\ifmmode\ddot{o}\else\"{o}\fi}n}}]{Schriefl2006Jan}%
  \BibitemOpen
  \bibfield  {author} {\bibinfo {author} {\bibfnamefont {J.}~\bibnamefont {Schriefl}}, \bibinfo {author} {\bibfnamefont {Y.}~\bibnamefont {Makhlin}}, \bibinfo {author} {\bibfnamefont {A.}~\bibnamefont {Shnirman}},\ and\ \bibinfo {author} {\bibfnamefont {G.}~\bibnamefont {Sch{\ifmmode\ddot{o}\else\"{o}\fi}n}},\ }\bibfield  {title} {\bibinfo {title} {{Decoherence from ensembles of two-level fluctuators}},\ }\href {https://doi.org/10.1088/1367-2630/8/1/001} {\bibfield  {journal} {\bibinfo  {journal} {New J. Phys.}\ }\textbf {\bibinfo {volume} {8}},\ \bibinfo {pages} {1} (\bibinfo {year} {2006})}\BibitemShut {NoStop}%
\bibitem [{\citenamefont {Lanting}\ \emph {et~al.}(2009)\citenamefont {Lanting}, \citenamefont {Berkley}, \citenamefont {Bumble}, \citenamefont {Bunyk}, \citenamefont {Fung}, \citenamefont {Johansson}, \citenamefont {Kaul}, \citenamefont {Kleinsasser}, \citenamefont {Ladizinsky}, \citenamefont {Maibaum}, \citenamefont {Harris}, \citenamefont {Johnson}, \citenamefont {Tolkacheva},\ and\ \citenamefont {Amin}}]{Lanting2009Feb}%
  \BibitemOpen
  \bibfield  {author} {\bibinfo {author} {\bibfnamefont {T.}~\bibnamefont {Lanting}}, \bibinfo {author} {\bibfnamefont {A.~J.}\ \bibnamefont {Berkley}}, \bibinfo {author} {\bibfnamefont {B.}~\bibnamefont {Bumble}}, \bibinfo {author} {\bibfnamefont {P.}~\bibnamefont {Bunyk}}, \bibinfo {author} {\bibfnamefont {A.}~\bibnamefont {Fung}}, \bibinfo {author} {\bibfnamefont {J.}~\bibnamefont {Johansson}}, \bibinfo {author} {\bibfnamefont {A.}~\bibnamefont {Kaul}}, \bibinfo {author} {\bibfnamefont {A.}~\bibnamefont {Kleinsasser}}, \bibinfo {author} {\bibfnamefont {E.}~\bibnamefont {Ladizinsky}}, \bibinfo {author} {\bibfnamefont {F.}~\bibnamefont {Maibaum}}, \bibinfo {author} {\bibfnamefont {R.}~\bibnamefont {Harris}}, \bibinfo {author} {\bibfnamefont {M.~W.}\ \bibnamefont {Johnson}}, \bibinfo {author} {\bibfnamefont {E.}~\bibnamefont {Tolkacheva}},\ and\ \bibinfo {author} {\bibfnamefont {M.~H.~S.}\ \bibnamefont {Amin}},\ }\bibfield  {title} {\bibinfo {title} {{Geometrical dependence of the low-frequency noise in
  superconducting flux qubits}},\ }\href {https://doi.org/10.1103/PhysRevB.79.060509} {\bibfield  {journal} {\bibinfo  {journal} {Phys. Rev. B}\ }\textbf {\bibinfo {volume} {79}},\ \bibinfo {pages} {060509} (\bibinfo {year} {2009})}\BibitemShut {NoStop}%
\bibitem [{\citenamefont {Quintana}\ \emph {et~al.}(2017)\citenamefont {Quintana}, \citenamefont {Chen}, \citenamefont {Sank}, \citenamefont {Petukhov}, \citenamefont {White}, \citenamefont {Kafri}, \citenamefont {Chiaro}, \citenamefont {Megrant}, \citenamefont {Barends}, \citenamefont {Campbell}, \citenamefont {Chen}, \citenamefont {Dunsworth}, \citenamefont {Fowler}, \citenamefont {Graff}, \citenamefont {Jeffrey}, \citenamefont {Kelly}, \citenamefont {Lucero}, \citenamefont {Mutus}, \citenamefont {Neeley}, \citenamefont {Neill}, \citenamefont {O{'}Malley}, \citenamefont {Roushan}, \citenamefont {Shabani}, \citenamefont {Smelyanskiy}, \citenamefont {Vainsencher}, \citenamefont {Wenner}, \citenamefont {Neven},\ and\ \citenamefont {Martinis}}]{Quintana2017Jan}%
  \BibitemOpen
  \bibfield  {author} {\bibinfo {author} {\bibfnamefont {C.~M.}\ \bibnamefont {Quintana}}, \bibinfo {author} {\bibfnamefont {Y.}~\bibnamefont {Chen}}, \bibinfo {author} {\bibfnamefont {D.}~\bibnamefont {Sank}}, \bibinfo {author} {\bibfnamefont {A.~G.}\ \bibnamefont {Petukhov}}, \bibinfo {author} {\bibfnamefont {T.~C.}\ \bibnamefont {White}}, \bibinfo {author} {\bibfnamefont {D.}~\bibnamefont {Kafri}}, \bibinfo {author} {\bibfnamefont {B.}~\bibnamefont {Chiaro}}, \bibinfo {author} {\bibfnamefont {A.}~\bibnamefont {Megrant}}, \bibinfo {author} {\bibfnamefont {R.}~\bibnamefont {Barends}}, \bibinfo {author} {\bibfnamefont {B.}~\bibnamefont {Campbell}}, \bibinfo {author} {\bibfnamefont {Z.}~\bibnamefont {Chen}}, \bibinfo {author} {\bibfnamefont {A.}~\bibnamefont {Dunsworth}}, \bibinfo {author} {\bibfnamefont {A.~G.}\ \bibnamefont {Fowler}}, \bibinfo {author} {\bibfnamefont {R.}~\bibnamefont {Graff}}, \bibinfo {author} {\bibfnamefont {E.}~\bibnamefont {Jeffrey}}, \bibinfo {author} {\bibfnamefont {J.}~\bibnamefont
  {Kelly}}, \bibinfo {author} {\bibfnamefont {E.}~\bibnamefont {Lucero}}, \bibinfo {author} {\bibfnamefont {J.~Y.}\ \bibnamefont {Mutus}}, \bibinfo {author} {\bibfnamefont {M.}~\bibnamefont {Neeley}}, \bibinfo {author} {\bibfnamefont {C.}~\bibnamefont {Neill}}, \bibinfo {author} {\bibfnamefont {P.~J.~J.}\ \bibnamefont {O{'}Malley}}, \bibinfo {author} {\bibfnamefont {P.}~\bibnamefont {Roushan}}, \bibinfo {author} {\bibfnamefont {A.}~\bibnamefont {Shabani}}, \bibinfo {author} {\bibfnamefont {V.~N.}\ \bibnamefont {Smelyanskiy}}, \bibinfo {author} {\bibfnamefont {A.}~\bibnamefont {Vainsencher}}, \bibinfo {author} {\bibfnamefont {J.}~\bibnamefont {Wenner}}, \bibinfo {author} {\bibfnamefont {H.}~\bibnamefont {Neven}},\ and\ \bibinfo {author} {\bibfnamefont {J.~M.}\ \bibnamefont {Martinis}},\ }\bibfield  {title} {\bibinfo {title} {{Observation of Classical-Quantum Crossover of $1/f$ Flux Noise and Its Paramagnetic Temperature Dependence}},\ }\href {https://doi.org/10.1103/PhysRevLett.118.057702} {\bibfield
  {journal} {\bibinfo  {journal} {Phys. Rev. Lett.}\ }\textbf {\bibinfo {volume} {118}},\ \bibinfo {pages} {057702} (\bibinfo {year} {2017})}\BibitemShut {NoStop}%
\bibitem [{\citenamefont {Gao}\ \emph {et~al.}(2025)\citenamefont {Gao}, \citenamefont {Wu}, \citenamefont {Sun}, \citenamefont {Chen}, \citenamefont {Deng}, \citenamefont {Ma}, \citenamefont {Miao}, \citenamefont {Song}, \citenamefont {Wan}, \citenamefont {Wang}, \citenamefont {Xia}, \citenamefont {Ying}, \citenamefont {Zhang}, \citenamefont {Shi}, \citenamefont {Zhao},\ and\ \citenamefont {Deng}}]{Gao2025Apr}%
  \BibitemOpen
  \bibfield  {author} {\bibinfo {author} {\bibfnamefont {R.}~\bibnamefont {Gao}}, \bibinfo {author} {\bibfnamefont {F.}~\bibnamefont {Wu}}, \bibinfo {author} {\bibfnamefont {H.}~\bibnamefont {Sun}}, \bibinfo {author} {\bibfnamefont {J.}~\bibnamefont {Chen}}, \bibinfo {author} {\bibfnamefont {H.}~\bibnamefont {Deng}}, \bibinfo {author} {\bibfnamefont {X.}~\bibnamefont {Ma}}, \bibinfo {author} {\bibfnamefont {X.}~\bibnamefont {Miao}}, \bibinfo {author} {\bibfnamefont {Z.}~\bibnamefont {Song}}, \bibinfo {author} {\bibfnamefont {X.}~\bibnamefont {Wan}}, \bibinfo {author} {\bibfnamefont {F.}~\bibnamefont {Wang}}, \bibinfo {author} {\bibfnamefont {T.}~\bibnamefont {Xia}}, \bibinfo {author} {\bibfnamefont {M.}~\bibnamefont {Ying}}, \bibinfo {author} {\bibfnamefont {C.}~\bibnamefont {Zhang}}, \bibinfo {author} {\bibfnamefont {Y.}~\bibnamefont {Shi}}, \bibinfo {author} {\bibfnamefont {H.-H.}\ \bibnamefont {Zhao}},\ and\ \bibinfo {author} {\bibfnamefont {C.}~\bibnamefont {Deng}},\ }\bibfield  {title} {\bibinfo {title}
  {{The effects of disorder in superconducting materials on qubit coherence}},\ }\href {https://doi.org/10.1038/s41467-025-58745-y} {\bibfield  {journal} {\bibinfo  {journal} {Nat. Commun.}\ }\textbf {\bibinfo {volume} {16}},\ \bibinfo {pages} {1} (\bibinfo {year} {2025})}\BibitemShut {NoStop}%
\bibitem [{\citenamefont {Kogan}(1996)}]{Kogan1996Aug}%
  \BibitemOpen
  \bibfield  {author} {\bibinfo {author} {\bibfnamefont {{\relax Sh}.}~\bibnamefont {Kogan}},\ }\href {https://doi.org/10.1017/CBO9780511551666} {\emph {\bibinfo {title} {{Electronic Noise and Fluctuations in Solids}}}}\ (\bibinfo  {publisher} {Cambridge University Press},\ \bibinfo {address} {Cambridge, England, UK},\ \bibinfo {year} {1996})\BibitemShut {NoStop}%
\bibitem [{\citenamefont {de~Graaf}\ \emph {et~al.}(2020)\citenamefont {de~Graaf}, \citenamefont {Faoro}, \citenamefont {Ioffe}, \citenamefont {Mahashabde}, \citenamefont {Burnett}, \citenamefont {Lindstr{\ifmmode\ddot{o}\else\"{o}\fi}m}, \citenamefont {Kubatkin}, \citenamefont {Danilov},\ and\ \citenamefont {Tzalenchuk}}]{deGraaf2020Dec}%
  \BibitemOpen
  \bibfield  {author} {\bibinfo {author} {\bibfnamefont {S.~E.}\ \bibnamefont {de~Graaf}}, \bibinfo {author} {\bibfnamefont {L.}~\bibnamefont {Faoro}}, \bibinfo {author} {\bibfnamefont {L.~B.}\ \bibnamefont {Ioffe}}, \bibinfo {author} {\bibfnamefont {S.}~\bibnamefont {Mahashabde}}, \bibinfo {author} {\bibfnamefont {J.~J.}\ \bibnamefont {Burnett}}, \bibinfo {author} {\bibfnamefont {T.}~\bibnamefont {Lindstr{\ifmmode\ddot{o}\else\"{o}\fi}m}}, \bibinfo {author} {\bibfnamefont {S.~E.}\ \bibnamefont {Kubatkin}}, \bibinfo {author} {\bibfnamefont {A.~V.}\ \bibnamefont {Danilov}},\ and\ \bibinfo {author} {\bibfnamefont {A.~{\relax Ya}.}\ \bibnamefont {Tzalenchuk}},\ }\bibfield  {title} {\bibinfo {title} {{Two-level systems in superconducting quantum devices due to trapped quasiparticles}},\ }\bibfield  {journal} {\bibinfo  {journal} {Sci. Adv.}\ }\textbf {\bibinfo {volume} {6}},\ \href {https://doi.org/10.1126/sciadv.abc5055} {10.1126/sciadv.abc5055} (\bibinfo {year} {2020})\BibitemShut {NoStop}%
\bibitem [{\citenamefont {G{\ifmmode\ddot{u}\else\"{u}\fi}nzler}\ \emph {et~al.}(2025{\natexlab{a}})\citenamefont {G{\ifmmode\ddot{u}\else\"{u}\fi}nzler}, \citenamefont {Rieger}, \citenamefont {Spiecker}, \citenamefont {Koch}, \citenamefont {Timco}, \citenamefont {Winpenny}, \citenamefont {Pop},\ and\ \citenamefont {Wernsdorfer}}]{Gunzler2025Feb}%
  \BibitemOpen
  \bibfield  {author} {\bibinfo {author} {\bibfnamefont {S.}~\bibnamefont {G{\ifmmode\ddot{u}\else\"{u}\fi}nzler}}, \bibinfo {author} {\bibfnamefont {D.}~\bibnamefont {Rieger}}, \bibinfo {author} {\bibfnamefont {M.}~\bibnamefont {Spiecker}}, \bibinfo {author} {\bibfnamefont {T.}~\bibnamefont {Koch}}, \bibinfo {author} {\bibfnamefont {G.~A.}\ \bibnamefont {Timco}}, \bibinfo {author} {\bibfnamefont {R.~E.~P.}\ \bibnamefont {Winpenny}}, \bibinfo {author} {\bibfnamefont {I.~M.}\ \bibnamefont {Pop}},\ and\ \bibinfo {author} {\bibfnamefont {W.}~\bibnamefont {Wernsdorfer}},\ }\bibfield  {title} {\bibinfo {title} {{Kinetic inductance coupling for circuit QED with spins}},\ }\bibfield  {journal} {\bibinfo  {journal} {arXiv}\ }\href {https://doi.org/10.48550/arXiv.2502.07605} {10.48550/arXiv.2502.07605} (\bibinfo {year} {2025}{\natexlab{a}}),\ \Eprint {https://arxiv.org/abs/2502.07605} {2502.07605} \BibitemShut {NoStop}%
\bibitem [{\citenamefont {G{\ifmmode\ddot{u}\else\"{u}\fi}nzler}\ \emph {et~al.}(2025{\natexlab{b}})\citenamefont {G{\ifmmode\ddot{u}\else\"{u}\fi}nzler}, \citenamefont {Janic}, \citenamefont {Rieger}, \citenamefont {Gosling}, \citenamefont {Zapata~Gonzalez}, \citenamefont {Field}, \citenamefont {Geisert}, \citenamefont {Bacher}, \citenamefont {Hohmann}, \citenamefont {Spiecker}, \citenamefont {Wernsdorfer},\ and\ \citenamefont {Pop}}]{Gunzler2025Oct}%
  \BibitemOpen
  \bibfield  {author} {\bibinfo {author} {\bibfnamefont {S.}~\bibnamefont {G{\ifmmode\ddot{u}\else\"{u}\fi}nzler}}, \bibinfo {author} {\bibfnamefont {B.}~\bibnamefont {Janic}}, \bibinfo {author} {\bibfnamefont {D.}~\bibnamefont {Rieger}}, \bibinfo {author} {\bibfnamefont {N.}~\bibnamefont {Gosling}}, \bibinfo {author} {\bibfnamefont {N.}~\bibnamefont {Zapata~Gonzalez}}, \bibinfo {author} {\bibfnamefont {M.}~\bibnamefont {Field}}, \bibinfo {author} {\bibfnamefont {S.}~\bibnamefont {Geisert}}, \bibinfo {author} {\bibfnamefont {A.}~\bibnamefont {Bacher}}, \bibinfo {author} {\bibfnamefont {J.}~\bibnamefont {Hohmann}}, \bibinfo {author} {\bibfnamefont {M.}~\bibnamefont {Spiecker}}, \bibinfo {author} {\bibfnamefont {W.}~\bibnamefont {Wernsdorfer}},\ and\ \bibinfo {author} {\bibfnamefont {I.}~\bibnamefont {Pop}},\ }\bibfield  {title} {\bibinfo {title} {{Data and code for the article "Spin Environment of a Superconducting Qubit in High Magnetic Fields"}},\ }\bibfield  {journal} {\bibinfo  {journal} {Zenodo}\ }\href
  {https://doi.org/10.5281/zenodo.17312111} {10.5281/zenodo.17312111} (\bibinfo {year} {2025}{\natexlab{b}})\BibitemShut {NoStop}%
\bibitem [{\citenamefont {Rieger}\ \emph {et~al.}(2023{\natexlab{b}})\citenamefont {Rieger}, \citenamefont {G{\ifmmode\ddot{u}\else\"{u}\fi}nzler}, \citenamefont {Spiecker}, \citenamefont {Nambisan}, \citenamefont {Wernsdorfer},\ and\ \citenamefont {Pop}}]{Rieger2023Jul}%
  \BibitemOpen
  \bibfield  {author} {\bibinfo {author} {\bibfnamefont {D.}~\bibnamefont {Rieger}}, \bibinfo {author} {\bibfnamefont {S.}~\bibnamefont {G{\ifmmode\ddot{u}\else\"{u}\fi}nzler}}, \bibinfo {author} {\bibfnamefont {M.}~\bibnamefont {Spiecker}}, \bibinfo {author} {\bibfnamefont {A.}~\bibnamefont {Nambisan}}, \bibinfo {author} {\bibfnamefont {W.}~\bibnamefont {Wernsdorfer}},\ and\ \bibinfo {author} {\bibfnamefont {I.~M.}\ \bibnamefont {Pop}},\ }\bibfield  {title} {\bibinfo {title} {{Fano Interference in Microwave Resonator Measurements}},\ }\href {https://doi.org/10.1103/PhysRevApplied.20.014059} {\bibfield  {journal} {\bibinfo  {journal} {Phys. Rev. Appl.}\ }\textbf {\bibinfo {volume} {20}},\ \bibinfo {pages} {014059} (\bibinfo {year} {2023}{\natexlab{b}})}\BibitemShut {NoStop}%
\end{thebibliography}%

\balancecolsandclearpage

\onecolumngrid
\section*{Appendices}
\vspace{0.6cm}
\twocolumngrid
\appendix

\section{Fabrication Details}
\label{sec:supp:fabrication}
The sample analyzed in this manuscript is fabricated on a double-side polished c-plane sapphire substrate using lift-off electron-beam lithography. 
A single resist layer of PMMA A4, coated with an \qty{8}{\nano\metre} aluminum anti-static layer, is patterned with a \qty{100}{\kilo\electronvolt} electron-beam writer. 
After patterning, the anti-static layer is removed using MF319 developer, which contains tetramethylammonium hydroxide (TMAH), followed by development of the PMMA resist in a \qty{6}{\celsius} isopropyl alcohol (IPA)/H$_2$O solution (1:3 volume ratio). 
Prior to metal deposition, the substrate undergoes a \qty{15}{\second} Ar/O$_2$ plasma cleaning process using a Kaufman ion source.
A \qty{20}{\nano\metre} granular aluminum layer is then deposited in a single evaporation step at room temperature, using an oxygen atmosphere at a chamber pressure of $\sim\qty{1e-4}{\milli\bar}$ and a deposition rate of approximately \qty{1}{\nano\metre\per\second}. 
A titanium gettering step is performed beforehand to enhance vacuum quality to $\sim\qty{1e-8}{\milli\bar}$ before evaporation.
In the lift-off process, the sample is sequentially submerged in an acetone bath, a \qty{30}{\minute} N-ethyl-2-pyrrolidone (NEP) bath with ultrasonic cleaning, and finally an ethanol bath. The final sample has a sheet resistance of $\qty{1}{\kilo\ohm/\sq}$.

\section{Sample Holder and Qubit Measurement}
\label{sec:supp:sampleholder}
\begin{figure}[tb]
\centering
\includegraphics[width=(.5\textwidth)]{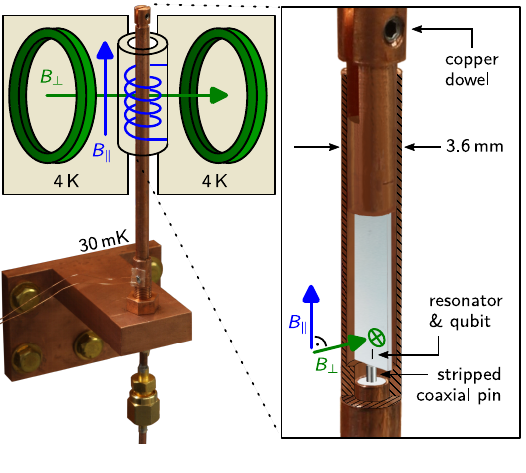}
\label{fig:SampleHolder}
\caption{
\textbf{Cylindrical waveguide sample holder within the vector magnet.} 
The 2D vector magnet is thermalized on the \qty{4}{\kelvin} stage of the cryostat and separated by a \qty{1}{\milli\metre} gap from the cylindrical pipe of the sample holder, which is anchored to the \qty{30}{\milli\kelvin} stage.
The waveguide, with a \qty{3}{\milli\metre} inner diameter and \qty{0,3}{\milli\metre} wall thickness, has a cut-off frequency of $\sim\qty{60}{\giga\hertz}$, operating in the sub-wavelength regime.
This results in coupling via the evanescent microwave field of a stripped coaxial pin, with an exponential decay in coupling strength relative to the chip-to-pin distance (cf.~\cite{Rieger2023Jul, gralmonium}).
The circuit is positioned on the bottom of a \qty{3}{\milli\metre} x \qty{10}{\milli\metre} sapphire chip and fixed with a copper dowel, clamped against the cylindrical copper pipe walls. 
Apiezon~N vacuum grease on the dowel provides additional thermal anchoring. 
High magnetic fields are applied in the substrate plane via a solenoid coil, while magnetic flux tuning is achieved with a Helmholtz pair aligned perpendicular to the substrate plane. 
No additional shielding is implemented between the sample holder and vector magnet coils.
Note that the setup is identical to the one used in Ref.\cite{Borisov2020Sep}
} 
\end{figure}
\begin{figure}[tb]
\centering
\includegraphics[width=(.5\textwidth)]{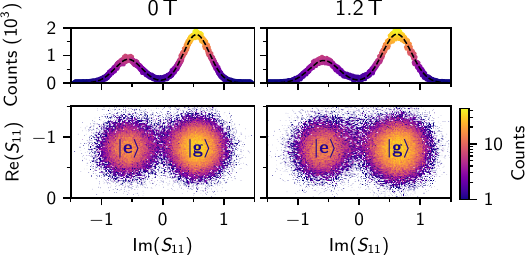}
\label{fig:IQhist}
\caption{
\textbf{IQ histogram in magnetic field:} 
1D histogram of the measured I quadrature (top panels) and 2D histogram of the I and Q quadrature (bottom panels).
In $B_\parallel=\qty{0}{\tesla}$ (left panels) and $B_\parallel=\qty{1,2}{\tesla}$ (right panels), we extract a qubit population corresponding to a temperature of \qty{165}{\milli\kelvin} and \qty{150}{\milli\kelvin}, respectively.
These IQ histograms are illustrative for all qubit state measurements used in the main text.
We use a \qty{540}{\nano\second} readout pulse with an equilibrium average photon number of $\bar{n} = 25$ , integrated over \qty{540}{\nano\second}, including resonator ring-up and ring-down times corresponding to a linewidth  $\kappa / 2\pi = \qty{1,2}{\mega\hertz}$.
} 
\end{figure}

In \cref{fig:SampleHolder}, we present the readout and thermalization of our sample housed within a copper waveguide sample holder (design identical to Refs.~\cite{Borisov2020Sep,gralmonium, Rieger2023Jul}).
The sample is thermally anchored to the \qty{30}{\milli\kelvin} stage of the cryostat by clamping it with a copper dowel into the sample holder and applying Apiezon~N vacuum grease for additional thermal contact.
The readout resonator, which we read out in single-port reflection,  is positioned in close proximity to the stripped pin of a coaxial cable to couple to its evanescent microwave field.
The 2D vector magnet is centered on the resonator and attached to the \qty{4}{\kelvin} stage of the cryostat.
We attribute the high thermal population of $p_\mathrm{th}\sim \qty{40}{\percent}$ at a qubit frequency of \qty{2,36}{\giga\hertz} (cf.~\cref{fig:hyperpolarization}) to thermal photons leaking into the sample holder due to the absence of additional thermal shielding.
In contrast, a similar device in the same sample holder geometry was reported in Ref.~\cite{gralmonium} with a qubit temperature of $\qty{37}{\milli\kelvin}$, where improved IR shielding and thermalization was possible due to the absence of the vector magnet.
In \cref{fig:IQhist}, we compare measured IQ histograms in $B_\parallel = \qty{0}{\tesla}$ and $B_\parallel = \qty{1,2}{\tesla}$ and we find similar qubit populations and signal to noise ratio.
To reduce thermal photon infiltration via the microwave lines, an infrared filter is placed in front of the sample holder.
From our setup we can identify potential culprits for increased magnetic field noise, such as vibrations of the cylindrical waveguide sample holder within the vector magnet, fluctuations in the vector magnet power supply or vortex retrapping within the coil.

To avoid a hysteretic, non-monotonic dependence of the resonator response on magnetic field, we apply high magnetic fields only in the substrate plane ($B_\parallel$) and minimize the out-of-plane component ($B_\perp$).
We account for minor chip misalignments, including tilt and rotation, by determining a compensation field $B_{\perp, \mathrm{comp}}$ for each $B_\parallel$, ensuring that all qubit measurements are conducted within the flux period closest to zero effective out-of-plane field.
Out-of-plane magnetic fields induce screening currents in the resonator that suppress the resonance frequency. 
We sweep $B_\perp$ and find the compensation field by maximizing the resonator frequency.
We determine an effective chip misalignment of \qty{0,66}{\milli\tesla/\tesla}.
A detailed description of this compensation procedure can be found in the supplementary information of Ref.~\cite{Borisov2020Sep}.

\section{Flux Sensitivity in Gradiometric and Non-Gradiometric Gralmonium Qubits}
\label{sec:supp:fluxsweep_comparison}
\begin{figure}[tb]
\centering
\includegraphics[width=(.5\textwidth)]{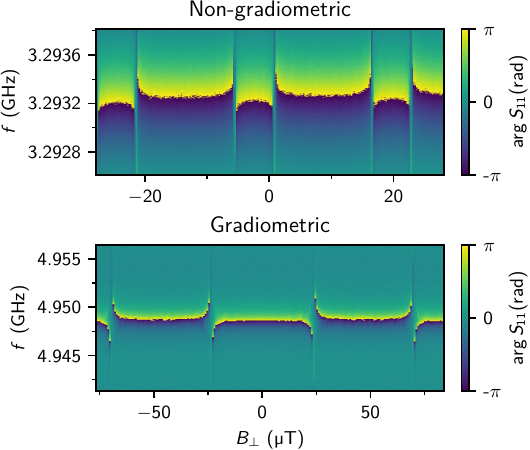}
\caption{
\textbf{Flux sensitivity in a gradiometric and non-gradiometric gralmonium.} 
A flux sweep of the resonator phase response $\mathrm{arg}(S_{11}) $ for non-gradiometric (top panel) and gradiometric (bottom panel) gralmonium qubits shows avoided level crossings where the qubit frequency intersects the resonator frequency. 
We observe a $4.6$-fold increase in flux periodicity from the non-gradiometric to the gradiometric qubit design.
} 
\label{fig:fluxsweep_comparison}
\end{figure}

In addition to the gradiometric qubit design shown in \crefadd{fig:sample}{a}, we fabricate a non-gradiometric fluxonium qubit with a similar layout:
Here, we removed the connecting wire that would close the second flux loop ($\Phi_{\mathrm{ext},2}$), resulting in a gralmonium with only one flux loop, illustrated in violet in \crefadd{fig:sample}{a} ($\Phi_{\mathrm{ext}}=\Phi_{\mathrm{ext},1}$, the inductance $L_3$ in \crefadd{fig:sample}{b} is removed).
In \cref{fig:fluxsweep_comparison}, we compare the flux periodicities of the gradiometric and non-gradiometric design, finding an increase by a factor of $4.6$.
This increase is consistent with the flux loop size ratio  $\Phi_{\mathrm{ext},1}/\Phi_\Delta$, suggesting negligible inductance asymmetry $\alpha\approx 0$.
Note that in our qubit implementation, one flux loop extends into the resonator, where the average screening current flows along the center of the  \qty{4}{\micro\metre}-wide strip due to the high impedance of the resonator, as illustrated in~\crefadd{fig:sample}{a}.

\section{Flux Noise in Magnetic Field}
\label{sec:supp:fluxNoise}
\begin{figure*}[tb]
\centering
\includegraphics{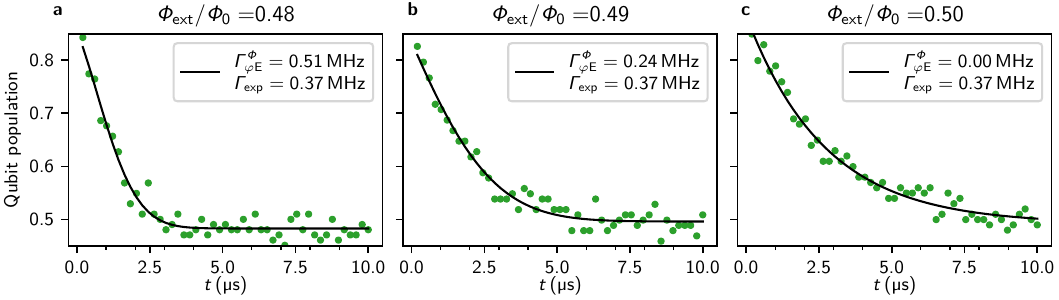}
\caption{
\textbf{Flux-dependence of the Hahn echo decay.} 
Echo measurement with a single refocusing $\pi$-pulse at three representative flux biases: \textbf{a}~$\Phi_\mathrm{ext}/\unit{\Phi_0}=0.48$, \textbf{b}~$\Phi_\mathrm{ext}/\unit{\Phi_0}=0.49$ and \textbf{c}~$\Phi_\mathrm{ext}/\unit{\Phi_0}=0.5$. 
 Black lines indicate a joint fit to $42$ individual echo measurements acquired over the flux range $\Phi_\mathrm{ext}=\qtyrange{0,48}{0,52}{\Phi_0}$, using~\cref{eq:Gaussian_Echo_fit}, which includes a constant exponential decay rate $\Gamma_\mathrm{exp}$ and a flux-dependent Gaussian dephasing contribution $\Gamma_{\varphi \text{E}}^\Phi$.
} 
\label{fig:Echo_Gaussian}
\end{figure*}
\begin{figure}[tb]
\centering
\includegraphics[width=(.5\textwidth)]{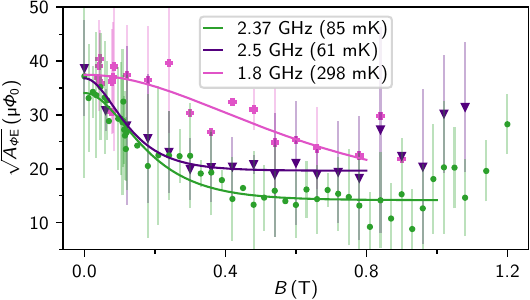}
\caption{
\textbf{Flux noise in magnetic field.} 
The flux noise amplitude from echo experiments (cf.~\cref{eq:GammaE_flux}) decreases in magnetic field. Different colors and marker shapes correspond to different qubit frequencies at half flux bias, obtained in different cooldowns of the same device.
Lines show fits to \cref{eq:APhi_vs_B}.
Marker shapes are identical to the inset in \crefadd{fig:coherence}{d}.
} 
\label{fig:fluxnoise}
\end{figure}

\Cref{fig:Echo_Gaussian} illustrates the fitting procedure for Hahn echo decay measurements as a function of external flux.
At the half-flux sweet spot ($\Phi_\mathrm{ext}/\unit{\Phi_0} = 0.5$), the decay is purely exponential, while deviations from this point ($\Phi_\mathrm{ext}/\unit{\Phi_0}\neq0.5$) reveal an additional Gaussian dephasing component, $\Gamma_{\varphi \text{E}}^\Phi$.
The echo decay is modeled as~\cite{Schriefl2006Jan}
\begin{align}
    P_{\ket{e}} (t) = \frac{1}{2}e^{-(\Gamma_{\varphi \text{E}}^\Phi t)^2} \cdot e^{-(\Gamma_1/2 + \Gamma_{\varphi\text{E}}^\text{const})t} + \frac{1}{2}\label{eq:Gaussian_Echo_fit}\,.
\end{align}
The exponential decay rate $\Gamma_\mathrm{exp} = \Gamma_1/2 + \Gamma_{\varphi\text{E}}^\text{const}$, combines energy relaxation and exponential dephasing, both assumed to be flux-independent over the range $\Phi_\mathrm{ext}=\qtyrange{0,48}{0,52}{\Phi_0}$.
Accordingly, we perform a joint fit for every set of Hahn echo measurements vs. flux,  extracting the flux-dependent Gaussian dephasing rate $\Gamma_{\varphi \text{E}}^\Phi$ atop the fixed exponential decay envelope defined by $\Gamma_\mathrm{exp}$.

In \cref{fig:fluxnoise}, we report the magnetic field dependence of the flux noise amplitude for three qubit frequencies at half flux bias. We obtain the change in the qubit frequency at half flux bias by thermal cycling of our device to room temperature.
We observe a suppression of the flux noise amplitude across all qubit frequencies, which we model using two-level fluctuators generating asymmetric random telegraph signals with corresponding Lorentzian power spectrum
\begin{align}
    S(\omega)\propto \frac{1}{\Gamma_1/\Gamma_\uparrow + \Gamma_1/\Gamma_\downarrow} \frac{\Gamma_1}{\Gamma_1^2 + \omega^2} \,.\label{eq:Lorentzian_PSD}
\end{align}
We assume the total decay rate $\Gamma_1$ remains invariant under magnetic field.
To rewrite the amplitude of the power spectrum in \cref{eq:Lorentzian_PSD} we use detailed balance $\dot{p_0} = \Gamma_\downarrow p_1 - \Gamma_\uparrow p_0$ where $p_0$ and $p_1$ are the population probabilities  for the ground and excited state, respectively.
At thermal equilibrium, $\dot{p_0}=0$, we obtain
\begin{align}
    &\Gamma_\uparrow = \Gamma_1 p_\mathrm{th}\\
    & \Gamma_\downarrow = \Gamma_1 (1-p_\mathrm{th})\,,
\end{align}
where $p_\mathrm{th}$ is the excited state population probability in thermal equilibrium.
Assuming Boltzmann-distributed populations
\begin{align}
    \frac{p_1}{p_0} = \frac{p_\mathrm{th}}{1-p_\mathrm{th}} =  e^{\frac{-\Delta E}{k_\mathrm{B} T}}\,,
\end{align}
with the energy difference $\Delta E$ between the ground and excited state, the flux noise power becomes
\begin{align}
    1/A_\Phi
    &\propto \frac{\Gamma_1}{\Gamma_\uparrow} + \frac{\Gamma_1}{\Gamma_\downarrow}\\
    &=  \frac{1}{  p_\mathrm{th}} + \frac{1}{(1-p_\mathrm{th})} \\
    &= \left(e^{\frac{-\Delta E}{k_\mathrm{B} T}}+1\right)
    \left(e^{\frac{\Delta E}{k_\mathrm{B} T}}+1\right)
    \\
    &= 4 \cosh^2\left(\frac{\Delta E}{2 k_\mathrm{B} T}\right)\\
    \sqrt{A_\Phi}&\propto \frac{1}{\cosh\left(\frac{\mu_\mathrm{B}B}{k_\mathrm{B} T}\right)}\,,\label{eq:sqrtAPhi_vs_B}
\end{align}
where we inserted the energy difference $2\mu_\mathrm{B}B$ of $g=2$ spin $s=1/2$ paramagnetic impurities in the last step.
Fits to \cref{eq:sqrtAPhi_vs_B} are in good agreement with the measured flux noise amplitudes (cf.~\cref{fig:fluxnoise}).
The observed variation in extracted spin bath temperatures motivates further noise characterization, including measurements at different cryostat temperatures.
To distinguish between various sources of flux-bias-dependent noise, such as flux noise from the spins, vortex noise in the coil, or kinetic inductance fluctuations in the qubit superinductor, a comprehensive analysis of the qubit power spectral density is required, similar to Ref.~\cite{Yan2016Nov}.

\section{TLS Hyperpolarization}
\label{sec:supp:TLShyperpol}
\begin{figure*}[t!]
\centering
\includegraphics[width=\textwidth]{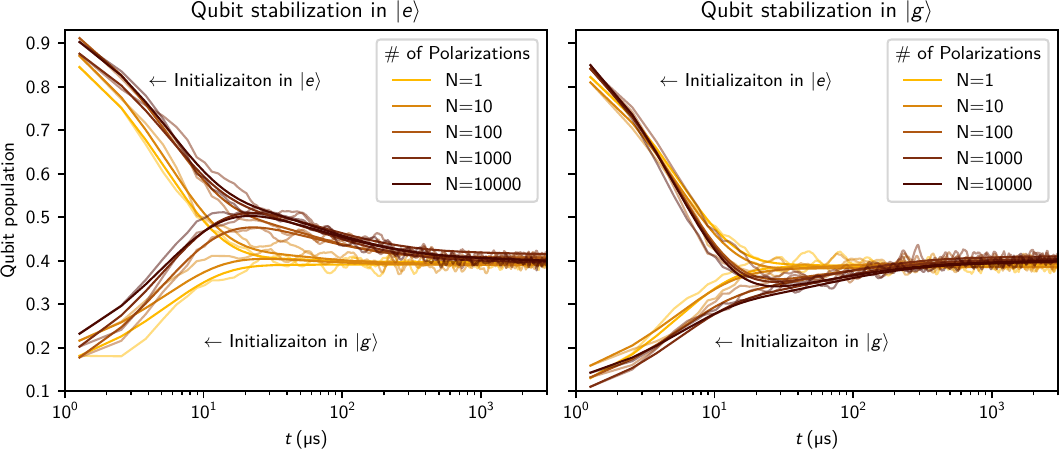}
\caption{
\textbf{TLS hyperpolarization experiment with varying numbers of qubit preparations in the stabilization sequence.} 
Qubit population relaxation,  measured stroboscopically at intervals of $\delta t=\qty{1.28}{\micro\second}$ following the preparation sequence shown in \crefadd{fig:hyperpolarization}{a}. 
The sequence consists of $N$ active state preparations in either $\ket{e}$~(left panel) or $\ket{g}$~(right panel), followed by qubit initialization in $\ket{g}$ or $\ket{e}$.
Semi-transparent lines show linearly sampled data on a logarithmic axis, smoothed using a Savitzky-Golay filter with a window length scaling with the square root of the data index. 
Opaque lines represent a joint fit of the full data set to the theoretical model described in Refs.~\cite{Spiecker2023Sep, Spiecker2024May} (see main text).
Note that the data for $N=10000$ is identical to that in \crefadd{fig:hyperpolarization}{b} in the main text.
} 
\label{fig:supp:hyperpolarization}
\end{figure*}

In \cref{fig:supp:hyperpolarization} we present the TLS hyperpolarization experiment for a varying number $N=\qtyrange{1}{e4}{}$  of active qubit state preparations within the stabilization sequence. 
We use an active qubit reset sequence, i.e. a readout followed by a conditional $\pi$-pulse which prepares the qubit in the target state ($\ket{e}$ or $\ket{g}$).
This feedback-based stabilization is repeated $N$ times (cf.~\crefadd{fig:hyperpolarization}{a}).
The cross-relaxation rate between the qubit and a TLS labeled by index $k$ is described by~\cite{Spiecker2023Sep, Spiecker2024May}
\begin{align}
    \Gamma_\mathrm{qt}^k = \frac{2g^2 \Gamma_2}{\Gamma_2^2 + \delta_k^2}\,, \label{eq:TLScrossrelaxation}
\end{align}
where $g$ is the qubit-TLS coupling strength, $\Gamma_2$ is the total decoherence rate of the coupled system, and $\delta_k$ is the frequency detuning between the qubit and the k-th TLS.
Through repeated stabilization, the TLSs are polarized towards the qubit stabilization state via the cross-relaxation (\cref{eq:TLScrossrelaxation}), resulting in significant population transfer. 
For stabilization in $\ket{e}$, this induces a population inversion (hyperpolarization) of the TLSs, exceeding the thermal limit of 50\%, similar to spectral hole burning.

To model the dynamics, following Ref.~\cite{Spiecker2023Sep} we consider a qubit coupled to a frequency-distributed ensemble of TLSs, governed by the Solomon rate equations~\cite{Spiecker2024May}:
\begin{align}
	\dot{p}_\text{q} &=  -\Gamma_\text{q} (p_\text{q} - p_\text{th}) - \sum_k\Gamma_\text{qt}^k (p_\text{q} - p_\text{t}^k)  \label{eq_rate_eq_qubit}\\
	\dot{p}_\text{t}^k &= - \Gamma_\text{t}^k (p_\text{t}^k - p_\text{th}) - \Gamma_\text{qt}^k (p_\text{t}^k - p_\text{q}), \label{eq_rate_eq_tls}
\end{align}
where $p_\text{q}$ and $p_\text{t}^k$ are the qubit and TLS populations, $\Gamma_\text{q}$ and $\Gamma_\text{t}^k$ are their respective intrinsic relaxation rates, and $p_\text{th}$ is the thermal equilibrium population.
To reduce model complexity, following Ref.~\cite{Spiecker2023Sep}, we assume TLSs are equally spaced in frequency, $\delta_k = k\Delta + \Delta_0$, with identical $g$ and $\Gamma_2$.
Here, $\Delta$ is the inter-TLS spacing, and $\Delta_0\in [0,\Delta/2]$ represents a frequency offset relative to the qubit.
The data in \cref{fig:supp:hyperpolarization} are jointly fitted using this simplified model, yielding a total cross-relaxation  $\sum \Gamma_\mathrm{qt}=\qty{45}{\kilo\hertz}$ and an intrinsic qubit relaxation rate $\Gamma_q=\qty{140}{\kilo\hertz}$.
\end{document}